\documentclass{emulateapj}
\usepackage[]{natbib, graphicx}

\shorttitle{Galaxy Zoo: AGN Host Galaxies}
\shortauthors{Schawinski et al.}
\slugcomment{Accepted for publication in the Astrophysical Journal}

\begin{document}

\title{Galaxy Zoo: The fundamentally different co-evolution of supermassive black holes and their early- and late-type host galaxies\altaffilmark{1}}

\author{
Kevin Schawinski,\altaffilmark{2,3,18}
C. Megan Urry,\altaffilmark{2,3,4}
Shanil Virani,\altaffilmark{3,4}
Paolo Coppi,\altaffilmark{3,4}
Steven P. Bamford,\altaffilmark{5}
Ezequiel Treister,\altaffilmark{6,19}
Chris J. Lintott,\altaffilmark{7}
Marc Sarzi,\altaffilmark{8}
William C. Keel,\altaffilmark{9}
Sugata Kaviraj,\altaffilmark{7,10}
Carolin N. Cardamone,\altaffilmark{3,4}
Karen L. Masters,\altaffilmark{11}
Nicholas P. Ross,\altaffilmark{12}
Dan Andreescu,\altaffilmark{13}
Phil Murray,\altaffilmark{14}
Robert C. Nichol,\altaffilmark{11}
M. Jordan Raddick,\altaffilmark{15}
An\v{z}e Slosar,\altaffilmark{16}
Alex S. Szalay,\altaffilmark{15}
Daniel Thomas\altaffilmark{11} and
Jan Vandenberg\altaffilmark{15}\\
}

\altaffiltext{1}{This publication has been made possible by the participation of more than 250,000 volunteers in the Galaxy Zoo project. Their contributions are individually acknowledged at
\texttt{http://www.galaxyzoo.org/Volunteers.aspx}.}
\altaffiltext{2}{Department of Physics, Yale University, New Haven, CT 06511, U.S.A.}
\altaffiltext{3}{Yale Center for Astronomy and Astrophysics, Yale University, P.O. Box 208121, New Haven, CT 06520, U.S.A.}
\altaffiltext{4}{Department of Astronomy, Yale University, New Haven, CT 06511, U.S.A.}
\altaffiltext{5}{Centre for Astronomy and Particle Theory, University of Nottingham, University Park, Nottingham, NG7 2RD, UK}
\altaffiltext{6}{Institute for Astronomy, 2680 Woodlawn Drive, University of Hawaii, Honolulu, HI 96822, U.S.A.}
\altaffiltext{7}{Department of Physics, University of Oxford, Keble Road, Oxford, OX1 3RH, UK}
\altaffiltext{8}{Centre for Astrophysics Research, University of Hertfordshire, College Lane, Hatfield, Herts AL10 9AB, UK}
\altaffiltext{9}{Department of Physics \& Astronomy, 206 Gallalee Hall, 514 University Blvd., University of Alabama, Tuscaloosa, AL 35487-0324, U.S.A.}
\altaffiltext{10}{Blackett Laboratory, Imperial College London, South Kensington Campus, London SW7 2AZ, UK}
\altaffiltext{11}{Institute of Cosmology and Gravitation, University of Portsmouth, Mercantile House, Hampshire Terrace, Portsmouth, PO1 2EG, UK}
\altaffiltext{12}{Department of Astronomy and Astrophysics, 525 Davey Laboratory, Pennsylvania State University, University Park, PA 16802.}
\altaffiltext{13}{LinkLab, 4506 Graystone Ave., Bronx, NY 10471, U.S.A.}
\altaffiltext{14}{Fingerprint Digital Media, 9 Victoria Close, Newtownards, Co. Down, Northern Ireland, BT23 7GY, UK}
\altaffiltext{15}{Department of Physics and Astronomy, The Johns Hopkins University, Homewood Campus, Baltimore, MD 21218, USA}
\altaffiltext{16}{Berkeley Center for Cosmological Physics, Lawrence Berkeley Nat. Lab \& Phys. Dept, University of California, Berkeley, CA 94720, U.S.A}
\altaffiltext{17}{Faculty of Mathematics \& Physics, University of Ljubljana, Slovenia}
\altaffiltext{18}{Einstein Fellow}
\altaffiltext{19}{Chandra Fellow}
\email{kevin.schawinski@yale.edu}

\def\Chandra{\textit{Chandra}}
\def\XMM{\textit{XMM-Newton}}
\def\Swift{\textit{Swift}}

\def\OI{[\mbox{O\,{\sc i}}]~$\lambda 6300$}
\def\OIII{[\mbox{O\,{\sc iii}}]~$\lambda 5007$}
\def\SII{[\mbox{S\,{\sc ii}}]~$\lambda \lambda 6717,6731$}
\def\NII{[\mbox{N\,{\sc ii}}]~$\lambda 6584$}

\def\Ha{{H$\alpha$}}
\def\Hb{{H$\beta$}}

\def\NIIHa{[\mbox{N\,{\sc ii}}]/H$\alpha$}
\def\SIIHa{[\mbox{S\,{\sc ii}}]/H$\alpha$}
\def\OIHa{[\mbox{O\,{\sc i}}]/H$\alpha$}
\def\OIIIHb{[\mbox{O\,{\sc iii}}]/H$\beta$}

\def\Ebmv{E($B-V$)}
\def\LOIII{$L[\mbox{O\,{\sc iii}}]$}
\def\Ledd{${L/L_{\rm Edd}}$}
\def\LOIIIs4{$L[\mbox{O\,{\sc iii}}]$/$\sigma^4$}
\def\LOIIIMbh{$L[\mbox{O\,{\sc iii}}]$/$M_{\rm BH}$}
\def\Mbh{$M_{\rm BH}$}
\def\Msigma{$M_{\rm BH} - \sigma$}
\def\Ms{$M_{\rm *}$}
\def\Msun{$M_{\odot}$}
\def\Msunyr{$M_{\odot}yr^{-1}$}

\def\ergs{$~\rm ergs^{-1}$}
\def\kms{$~\rm kms^{-1}$}

\begin{abstract}
We use data from the Sloan Digital Sky Survey and visual classifications of morphology from the Galaxy Zoo project to study black hole growth in the nearby Universe ($ z < 0.05$) and to break down the AGN host galaxy population by color, stellar mass and morphology. We find that black hole growth at luminosities \LOIII\ $>10^{40}$\ergs\ in early- and late-type galaxies is fundamentally different. AGN host galaxies as a population have a broad range of stellar masses ($10^{10} - 10^{11}$\Msun), reside in the green valley of the color-mass diagram and their central black holes have median masses around $10^{6.5}$\Msun. However, by comparing early- and late-type AGN host galaxies to their non-active counterparts, we find several key differences: in early-type galaxies, it is preferentially the galaxies with the \textit{least massive} black holes that are growing, while late-type galaxies, it is preferentially the \textit{most massive} black holes that are growing. The duty cycle of AGN in early-type galaxies is strongly peaked in the green valley below the low-mass end ($10^{10}$\Msun) of the red sequence at stellar masses where there is a steady supply of blue cloud progenitors. The duty cycle of AGN in late-type galaxies on the other hand peaks in massive ($10^{11}$\Msun) green and red late-types which generally do not have a corresponding blue cloud population of similar mass. At high Eddington ratios (\Ledd\ $>0.1$), the only population with a substantial fraction of AGN are the low-mass green valley early-type galaxies. Finally, the Milky Way likely resides in the ``sweet spot'' on the color-mass diagram where the AGN duty cycle of late-type galaxies is highest.  We discuss the implications of these results for our understanding of the role of AGN in the evolution of galaxies.
\end{abstract}


\section{Introduction}
The nature of the co-evolution between galaxies and their supermassive black holes is still a mystery
a decade after the discovery of the relationship between spheroid mass and black hole mass \citep{1998AJ....115.2285M,2000ApJ...539L..13G, 2000ApJ...539L...9F,2002ApJ...574..740T, 2004ApJ...604L..89H}. Even more mysterious is its connection to the phenomenon of \textit{downsizing}, the anti-hierarchical growth of galaxies and black holes. 
While dark matter grows and assembles in a strictly hierarchical  fashion, with the most massive dark matter haloes continuing to grow  over cosmic time by accretion of smaller haloes, the formation and  growth of galaxies inside those halos proceeds differently \citep[e.g.,][]{1977ApJ...211..638S, 1977MNRAS.179..541R, 1978MNRAS.183..341W, 1980MNRAS.193..189F, 1984Natur.311..517B, 1996MNRAS.281..716C} . Baryonic matter is converted into stars within dark matter haoles and regulated by feedback processes. The stars in the most massive galaxies were not only formed first, but also on the shortest time scales, with further star formation at later epochs being suppressed. Less massive galaxies have increasingly more extended star formation histories as star formation in them, and their progenitors, proceeded over increasingly longer timescales. This picture has been verified both by ``archeological studies'' of local galaxies, and by probing ongoing star formation at increasing redshifts \citep{1996AJ....112..839C, 2004Natur.428..625H, 2005ApJ...621..673T, 2005ApJ...632..137N, 2006AJ....131.1288B, 2006ApJ...651..120B, 2007ApJ...669..947J, 2008ApJ...675..234P, 2009Natur.458..603C, 2009arXiv0912.0259T}, and has alternatively been called `downsizing' or `anti-hierarchical' growth.

The growth of supermassive black holes seems to broadly mirror the anti-hierarchical trend of their host galaxies. Quasars powered by $\sim10^9$\Msun\ black holes are already observed about 1 Gyr after the Big Bang, implying that they must have reached these masses in a very brief, intense growth phase \cite[e.g.,][]{2001AJ....122.2833F, 2003AJ....125.1649F, 2004AJ....128..515F, 2006NewAR..50..665F}. Studies of the luminosity function of optical- and X-ray-selected active galactic nuclei (AGN) in deep fields observed by \Chandra\ and \XMM\ have revealed that the most luminous AGN were most common in the early Universe, while the lower-luminosity AGN peak later \citep{2003ApJ...598..886U, 2004MNRAS.353.1035M, 2005AJ....129..578B, 2005ApJ...635..864L,2009MNRAS.399.1755C}. In other words, the pattern of black hole growth inferred from the luminosity evolution of AGN seems to exhibit a qualitatively similar trend as galaxy growth. Studies attempting to recover the history of black hole growth  support this picture \citep{2004MNRAS.351..169M, 2009ApJ...694..867S}.

It has been suggested in recent years that energy input from AGN not only self-regulates the growth of both black holes and their host galaxies \citep{1998A&A...331L...1S, 2003ApJ...596L..27K} to give rise to the observed \Msigma\ relation  \citep{2000ApJ...539L..13G, 2000ApJ...539L...9F}, but that this \textit{AGN feedback} also, as \cite{2006MNRAS.370..645B} described it, ``breaks the hierarchy of galaxy formation'' giving rise to the observed anti-hierarchical nature of galaxy star formation histories. However, the precise nature of this connection is not understood, and it is not clear if all or only some AGN phases affect their host galaxies.

In this paper, we explore properties of the host galaxies of AGN and the nature of their black hole growth  in the local Universe ($z < 0.05$) and discuss what these properties imply for the role of AGN in the evolution of the host galaxies. In our investigation, we go beyond characterizing AGN host galaxies as a single class, as in some previous studies \citep[e.g.,][]{2003MNRAS.346.1055K, 2004ApJ...613..109H}, and dissect the AGN host galaxy population into sub-populations by observed properties such as mass, color and most importantly, morphology \citep[e.g.,][]{2005MNRAS.362...25B, 2009ApJ...696..891H}. We use the visual classifications of morphology by citizen scientists taking part in the Galaxy Zoo project  \citep{2008MNRAS.389.1179L} to divide both AGN host galaxies and their quiescent counterparts into morphology classes to reveal several new and surprising properties of AGN host galaxies that have implications for our picture of the galaxy-black hole co-evolution. Our main conclusion is that the early- and late-type host galaxies of AGN are, despite apparent similarities, very different populations, and that the AGN phases occurring in them play a very different role in their co-evolution.

In Section \ref{sec:sample}, we describe our sample selection and the derivation of galaxy properties such as stellar masses and morphologies. In Section \ref{sec:eml}, we motivate and describe the selection of AGN via emission line diagnostic diagrams and the derivation of accretion properties and black hole masses. We present our observational results in Section \ref{sec:results}, discuss their implications in Section \ref{sec:discussion} and summarize in Section \ref{sec:summary}. 

Throughout this work, we assume cosmological parameters $(\Omega_{\rm m} = 0.3, \Omega_{\Lambda} = 0.7, H_{0} = 70)$, consistent with the \textit{Wilkinson Microwave Anisotropy Probe} Third Year results in combination with other data \citep{2007ApJS..170..377S}. 

\section{Data and Sample Selection}
\label{sec:sample}

\begin{figure*}[!ht]
\begin{center}

\includegraphics[angle=90, width=\textwidth]{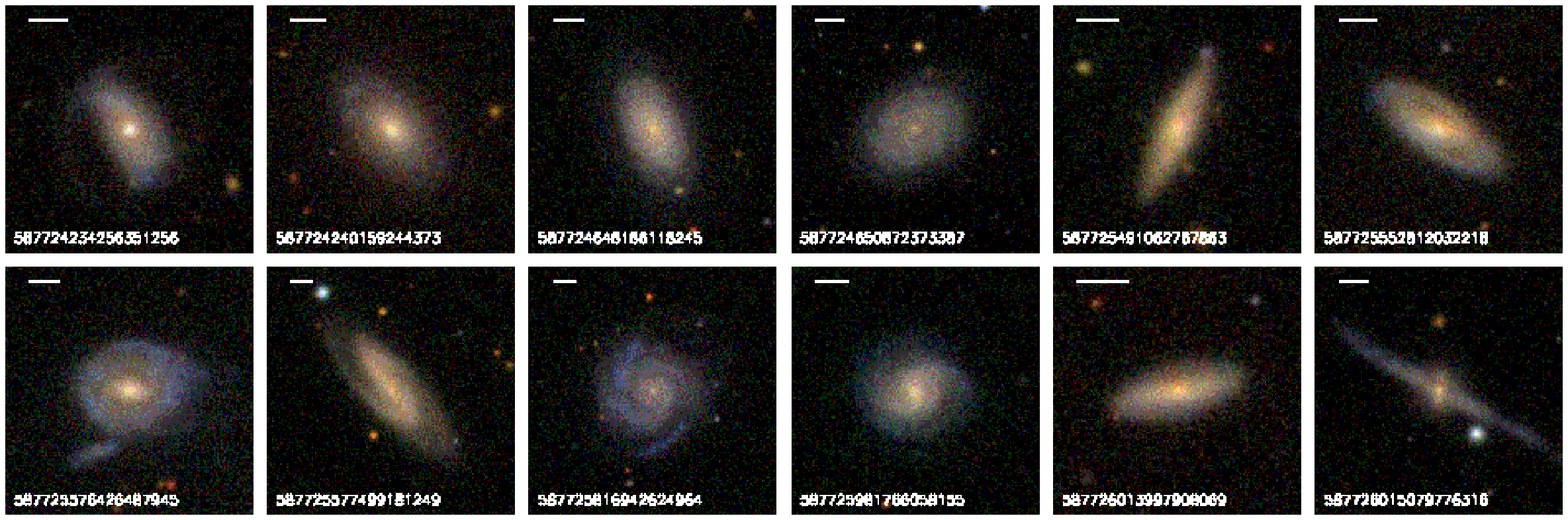}
\vspace{0.1in}\\
\includegraphics[angle=90, width=\textwidth]{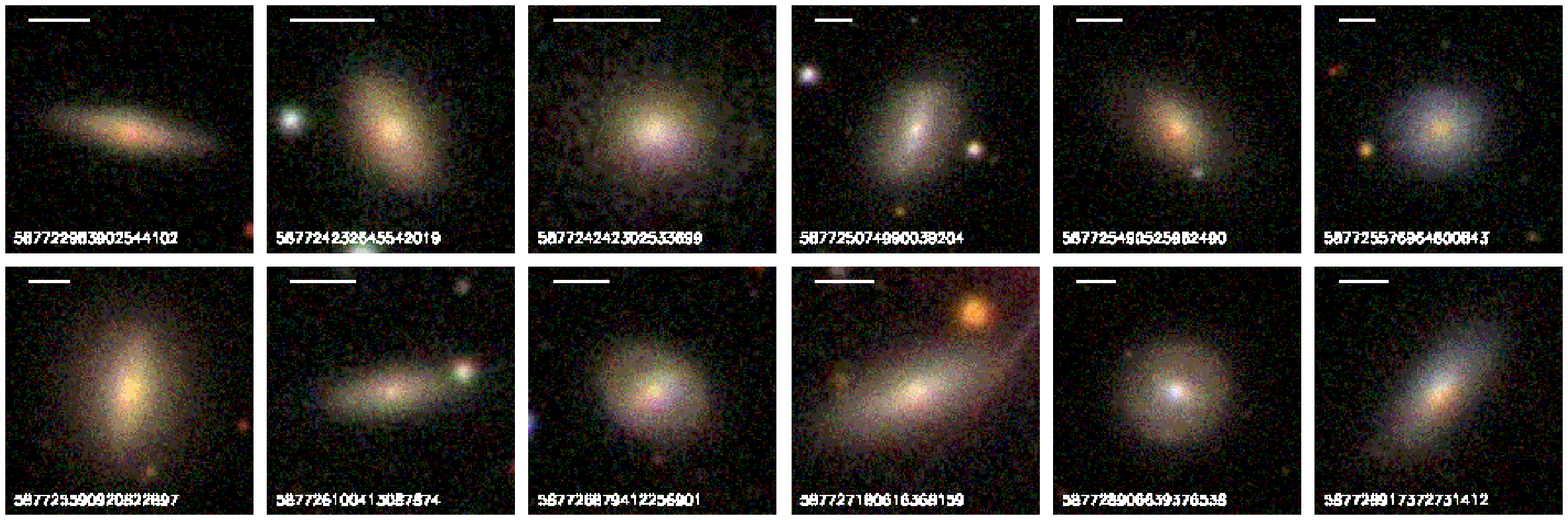}
\vspace{0.1in}\\
\includegraphics[angle=90, width=\textwidth]{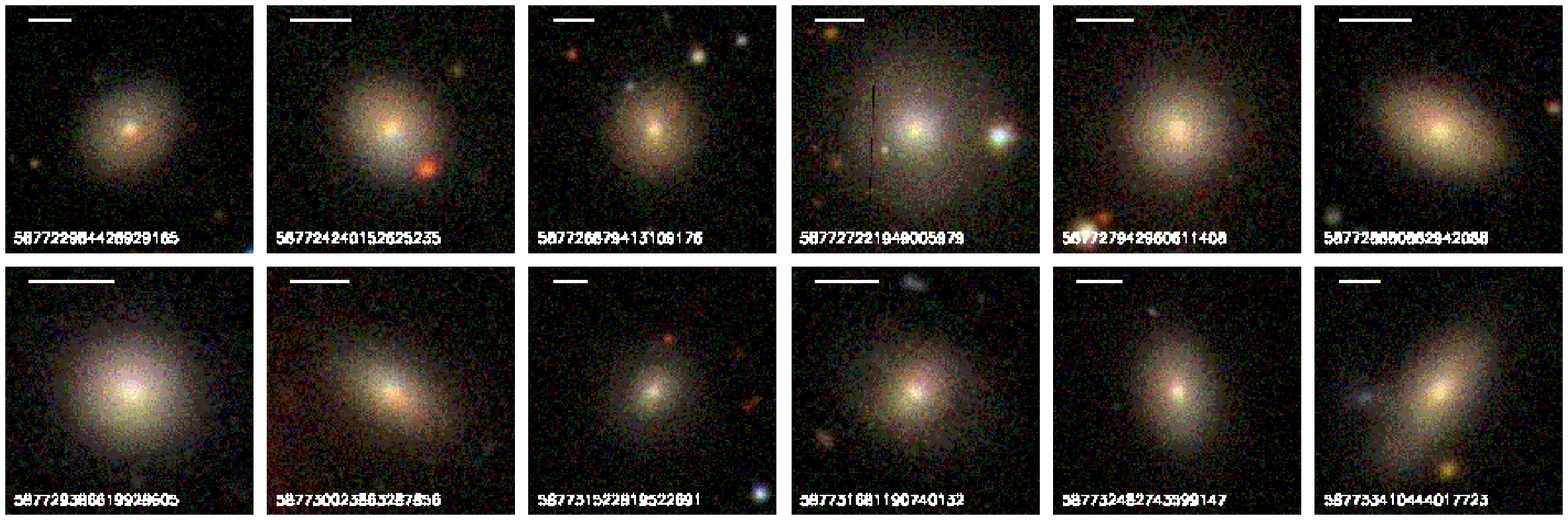}
\caption{Example images of AGN host galaxies in three morphology classes. The \textit{top} two rows are late-type host galaxies, the \textit{middle} rows are indeterminate-type and the \textit{bottom} two rows are early-type host galaxies. In the \textit{top left} of each image, we show a bar of 5 kpc length at the redshift of the object. In the bottom, we provide the SDSS object ID.\label{fig:sy_examples}}

\end{center}
\end{figure*}

In this section, we describe the selection of both the parent galaxy population and the AGN host galaxy sample we use in this paper. 

\subsection{SDSS Photometric and Spectroscopic Data}
The photometric and spectroscopic data used here are taken from the Sloan Digital Sky Survey (SDSS) DR7 \citep{2000AJ....120.1579Y, 2002AJ....124.1810S, 2009ApJS..182..543A}. Throughout the paper, we use \texttt{modelMag}s for galaxy colors, as they are derived from the best-fitting exponential or de Vaucouleurs galaxy profile and are thus best suited for measuring galaxy colors.

\subsection{Galaxy Sample Selection Criteria}
Our sample selection criteria ensure a large and complete sample of galaxies to low stellar masses whose SDSS images are suitable for detailed morphological classification. We select all galaxies with SDSS spectra classified as \texttt{GALAXY} \citep{2002AJ....124.1810S} in the redshift interval $0.02 < z < 0.05$ and limit our selection to galaxies with $r < 17$ AB mag. In order to create a magnitude-limited sample that is close to a stellar mass-limited sample, we further limit this sample to $M_{z, Petro} < -19.5$ AB mag. We choose the $z$-band rather than the $r$-band since the redder $z$-band is closer to stellar mass. This yields a total of 47,675 galaxies.

The SDSS fibres are circular with a diameter of 3\arcsec\ \citep{2000AJ....120.1579Y, 2002AJ....124.1810S}, which corresponds to a physical footprint of 1.2 -- 2.9 kpc in diameter at the lower and upper redshift limits of our sample. The spectroscopic information used in this work therefore samples only the central regions of galaxies, rather than the bulk light which is sampled in more distant ($z \sim 0.1$) galaxies.  We describe the selection of AGN from this parent sample in Section \ref{sec:eml}.

\subsection{Galaxy Zoo Morphologies}
\label{sec:gz_morph}

\begin{deluxetable}{lrr}[!h]
\tablecolumns{3}
\tablewidth{0pc}
\tabletypesize{\scriptsize}
\tablecaption{Galaxy morphology statistics from Galaxy Zoo}
\tablehead{
 \colhead{Galaxy morpholgy class} & 
 \colhead{Number} & 
 \colhead{Percentage}
}
\startdata
Late-type      		 &	16246	&        34.1\\
Indeterminate-type        &	22483    & 	   47.2\\
Early-type        		 &	8928      & 	  18.7\\
\hline
All				        &   47675     & 	100\\
\enddata
\label{tab:gzmorph}
\end{deluxetable}

In the past, obtaining accurate morphologies for large samples of galaxies has been difficult. Only with the involvement of over 200,000 citizen scientists in the Galaxy Zoo project  \citep{2008MNRAS.389.1179L} have we been able to gather multiple independent visual classifications for the entire SDSS \texttt{main} galaxy sample that are comparable to those of professional astronomers. Galaxy Zoo provides a web interface\footnote{See \texttt{http:// zoo1.galaxyzoo.org}} where volunteers from the general public are shown randomly selected images from the SDSS main galaxy sample. Since every object supplied for inspection by Galaxy Zoo has been viewed multiple times by independent classifiers, we are able to use the distribution of votes to choose the purity of the classifications of individual galaxies by specifying the required level of agreement among classifiers.

We use the definition for the \texttt{clean} sample of \cite{2008MNRAS.388.1686L}, which requires an 80\% majority agreement on the morphology of any object. We combine all spirals regardless of orientation into the \textit{late-type} class and all the spheroids into the \textit{early-type} class, while those objects without a consensus at the 80\% level are placed in the \textit{indeterminate} class. We use the label `early-type' rather than `elliptical', as we specifically include lenticular galaxies in this class, while the late-type class contains anything with a discernible disk, from an Sa to an Sd galaxy. 

We include merging galaxies with the indeterminate-type class; in any case, major mergers in the local Universe are rare \citep[1-2\%;][and references therein]{2009arXiv0903.4937D}, and the incidence of emission-line AGN in them is not significantly enhanced over the general population \citep{2009MNRAS.tmp.1770D}. In Table \ref{tab:gzmorph}, we present the total numbers and population fractions of the various morphology classes defined by the Galaxy Zoo classifications .

In order to give a sense what these three morphological classes mean, we show in Figure \ref{fig:sy_examples} a randomly selected sample of AGN host galaxies in each morphology class class. We stress that these are AGN host galaxies and not objects taken from the entire population. For examples of images of normal galaxies, we refer the reader to \cite{2008MNRAS.389.1179L}.

\subsection{Stellar Mass Measurements}
\label{sec:stellarmass}
We measure stellar masses for all objects in this sample by fitting the five SDSS photometric bands to a library of $6.8 \times 10^{6}$ model star formation histories generated from \cite{1998MNRAS.300..872M, 2005MNRAS.362..799M} stellar models. The star formation histories are parameterized as two bursts \citep[e.g.,][]{2007MNRAS.382.1415S, 2007ApJS..173..619K} of varying ages and mass-fractions, fixed solar metallicity, varying dust extinction ($0 < $\Ebmv$ < 0.5$; \citealt{2000ApJ...533..682C}) and exponential e-folding time ($\tau = 10, 100, 1000$ Myr); all models assume a \cite{1955ApJ...121..161S} initial mass function. Stellar masses are measured by finding the minimum of the $\chi^2$ statistic in the parameter space probed, while the errors are determined from the range of stellar masses within $1-\sigma$ of the minimum. We do not vary the initial mass function.

We test the reliability of our stellar mass measurements for AGN host galaxies with the methodology described in Appendix \ref{sec:reliability}. We find that our stellar mass measurements are robust for the AGN present in our sample.

\section{AGN Selection}
\label{sec:eml}

The selection and characterization of AGN is a challenging problem, as no single wavelength regime allows for a perfectly unbiased sample. Various mid-infrared selection techniques have been proposed based on \textit{Spitzer} data \citep[e.g.,][]{2004ApJS..154..166L,2005ApJ...631..163S, 2005ApJ...634..169D, 2009ApJ...696..891H}, but these can be inefficient depending on depth, missing a large fraction of genuine AGN while selecting a substantial number of non-AGN \citep{2008ApJ...680..130C}. Hard X-ray selection ($E > 10$ keV) is the least biased selection technique, as it can detect even the most heavily obscured, Compton-thick systems whose existence is required by the X-ray background  \citep{2005ApJ...630..115T,2009ApJ...696..110T}. However, current surveys of the hard X-ray sky remain too shallow or are confined to small areas compared to SDSS \citep[e.g.,][]{2008ApJ...681..113T, 2008ApJ...674..686W, 2009ApJ...696..110T}.

For this reason, we resort to a less complete, but more sensitive selection technique than hard X-rays by searching for galaxies exhibiting narrow emission lines originating from a cone of radiation photoionized by the central engine \citep[e.g.,][]{1991ApJ...369L..27E,1993ApJ...417...82E}. First used by \cite*{1981PASP...93....5B} and extended by \cite{1987ApJS...63..295V}, emission line diagrams have been improved and refined in their interpretation by many theorists using photoionisation codes \citep[e.g.,][]{1990A&AS...83..501S, 1995ApJS...96....9L, 1999ApJS..123....3L, 2001ApJ...556..121K, 2006MNRAS.372..961K, 2004ApJS..153....9G, 2004ApJS..153...75G} and by observational work \citep[e.g.,][]{1987ApJS...63..295V, 1995ApJS...98..477H, 1997ApJS..112..315H, 1997ApJ...487..568H, 1997ApJS..112..391H, 2003MNRAS.346.1055K, 2006MNRAS.371..972S, 2006MNRAS.366.1151S, 2008MNRAS.391L..29S}.

The emission line selection we use in this work does not include unobscured (Type 1) AGN with broad lines and may be biased against the most highly obscured AGN. If the Unified Model of AGN \citep{1993ARA&A..31..473A, 1995PASP..107..803U} is correct and these difference are due to the orientation of the putative obscuring medium in the central engine, and if the obscuration does not cover the narrow-line region, then narrow-line (Type 2) AGN are a fair sub-set of the entire AGN population within the luminosity range covered. 

We now discuss in detail how we measure emission lines, use them on emission line diagrams, and assess the completeness of this selection. The final result of our selection described in this Section is that of the high-luminosity end of the local AGN population, and therefore of sites of substantial black hole growth. For reasons of completeness, a similar analysis of low-luminosity AGN with significantly sub-Eddington accretion rates is not possible via emission line selection due to contamination from star formation (see Section \ref{sec:complete} for details).

\subsection{Emission Line Measurement}
\label{sec:linemeasure}

We measure the emission line fluxes using an analysis tool called Gas AND Absorption Line Fitting algorithm (\textsc{GANDALF};  \citealt{2006MNRAS.366.1151S}), which uses the Penalized Pixel-Fitting method \textsc{PPXF} method of \cite{2004PASP..116..138C} to simultaneously fit the stellar continuum and nebular emission lines\footnote{Both \textsc{PPXF} and \textsc{GANDALF} are available for download at \texttt{http://www.strw.leidenuniv.nl/sauron/software.html}. A version of \textsc{GANDALF} designed for use on SDSS spectra can be obtained at \texttt{http://star-www.herts.ac.uk/  $\sim$sarzi/PaperV\_nutshell/PaperV\_nutshell.html}}. The advantage of this tool is that it is designed to minimize template mismatch, where emission and absorption lines are difficult to separate (see \citealt{2006MNRAS.369..529F} and \citealt{2006MNRAS.366.1151S} for example applications). We use a set of stellar templates from \cite{2004ApJ...613..898T} plus Gaussian emission line templates. We measure \OIII\ fluxes for each processed SDSS spectrum and compute the total luminosity \LOIII\ by correcting for extinction based on the measured Balmer decrement. \textsc{GANDALF} also provides measurements of the stellar velocity dispersion. Since the fixed SDSS spectroscopic fibre samples the light out to a different fraction of the effective radius, we use the empirical correction derived by \cite{2006MNRAS.366.1126C} to correct the velocity dispersion to the effective radius of the bulge surface-brightness profile.

\subsection{Emission Line Diagnostics and AGN Selection}

\begin{figure*}[!ht]
\begin{center}

\includegraphics[angle=90, width=\textwidth]{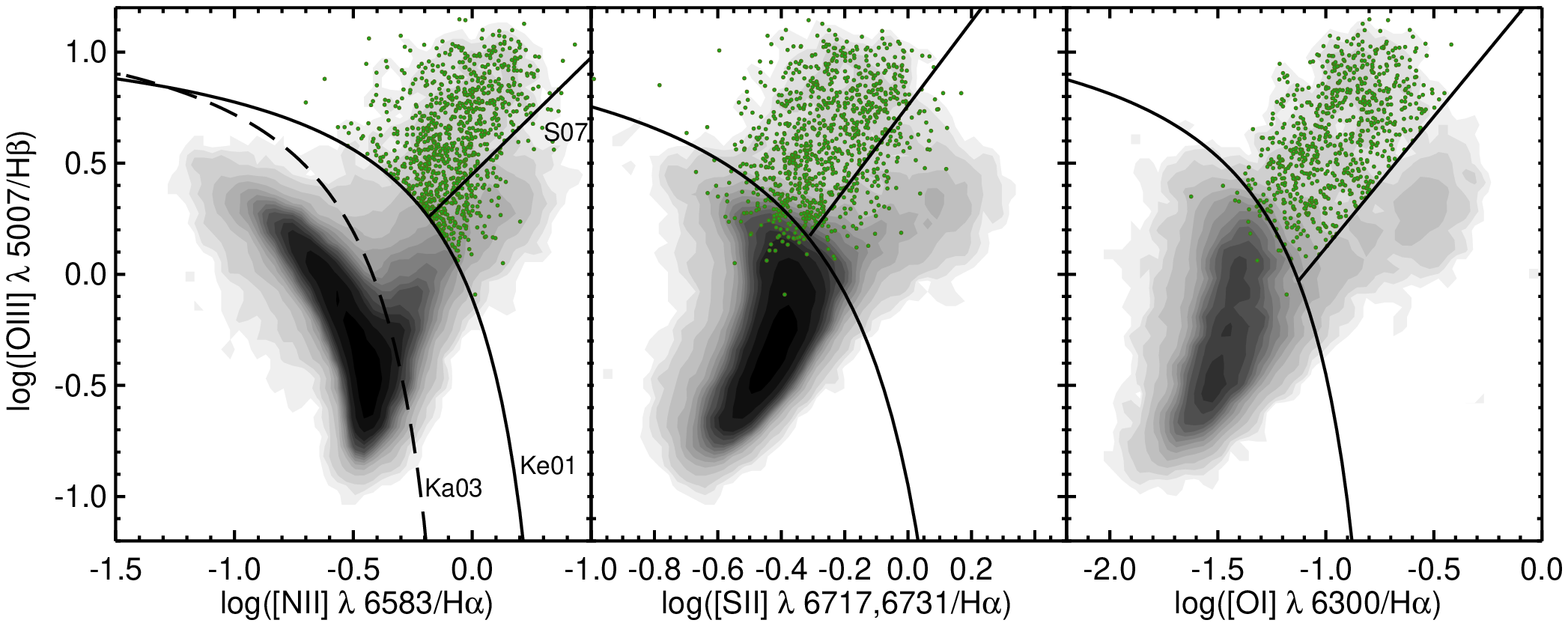}
\caption{The emission line diagnostic diagrams used for the selection of AGN from narrow-line fluxes. We use three different diagrams, all of which use the ratio of \OIIIHb\ (\textit{vertical axis}). From left to right, the \textit{horizontal axes} are the ratios of \NIIHa, \SIIHa\ and \OIHa. In each panel, we plot those objects that have S/N $>$ 3 in each of the four lines required for the diagram. The gray shading represent the entire galaxy sample, while the green points are the AGN sample selected using these diagrams. In the \NIIHa\ diagram (\textit{left}), the \textit{dashed line} shows the empirical demarcation line between between purely star-forming galaxies and the composite region of the diagram determined by \citet[][Ka03]{2003MNRAS.346.1055K}. The \textit{solid line}, labelled Ke01, is the theoretical extreme starburst line of \citet{2001ApJ...556..121K} beyond which the dominant source of ionization cannot be due to OB stars and must be due to a mechanism other than star formation (AGN, evolved stellar populations or shocks). The \textit{solid line} labelled S07 is the empirical AGN-LINER distinction defined by \cite{2007MNRAS.382.1415S}. The demarcation lines in the \SIIHa\ and \OIHa\ diagrams are taken from \citet{2006MNRAS.372..961K}. We select AGN using these diagrams in the following way: if \OI\ is detected with S/N $>3$, then we use the \OIHa\ diagram. If \OI\ is not detected, but \SII\ is, then we use the \SIIHa\ diagram. If neither \OI\ nor \SII\ are detected, but \NII\ is, we use the \NIIHa\ diagram. \label{fig:bpt}}

\includegraphics[angle=90, width=\textwidth]{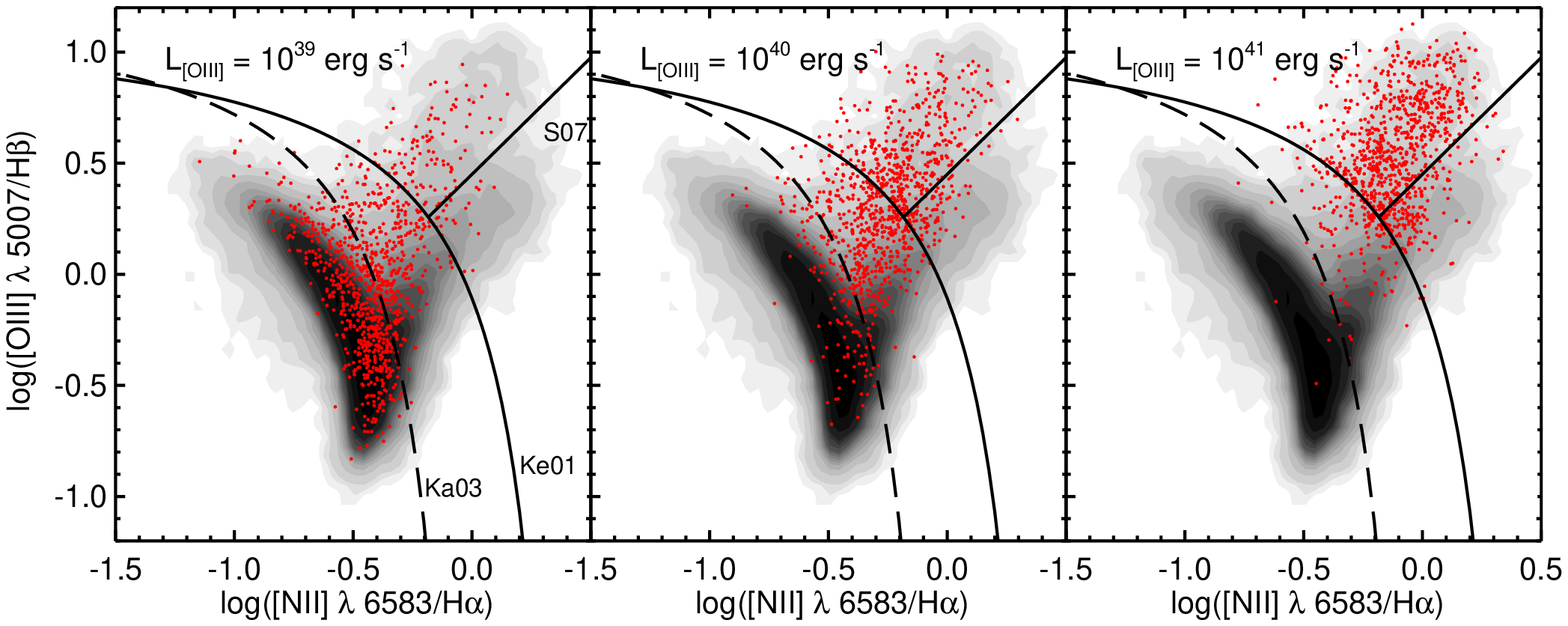}
\caption{\NIIHa\ emission line diagnostic diagrams for all galaxies (\textit{shaded}) and simulated composite AGN plus star forming galaxies (\textit{red points}). The simulated AGN luminosity increases from $10^{39}$\ergs in the \textit{left} panel, to $10^{40}$\ergs in the \textit{center} and $10^{41}$\ergs on the \textit{right}. See Section \ref{sec:complete} for details. For low values of \LOIII\ (\textit{left}), the nebular emission from star formation can overwhelm the AGN signature, and such low-luminosity AGN may be missed. At an AGN \LOIII = $10^{40}$\ergs,  12.8\% of AGN move below the empirical starburst line (\textit{dashed}), while 54.4\% move below the extreme starburst line (\textit{solid}). The result of this test shows that at high luminosities, AGN are efficiently selected using emission line diagrams. However, at lower luminosities, star formation may overwhelm the AGN narrow line emission. \label{fig:bpt2}}

\end{center}
\end{figure*}

Our AGN selection technique is the same as used by \cite{2007MNRAS.382.1415S}: we first separate galaxies whose narrow emission lines are dominated partially or wholly by star formation on the \NIIHa\ diagram (Figure \ref{fig:bpt}, \textit{left}) by using the extreme starburst line of \cite{2001ApJ...556..121K} to retain only those beyond this line (i.e., the top-right of the diagram). Between this extreme starburst line and the empirical pure starburst line of \cite{2003MNRAS.341...33K}, there is a substantial population of composite objects where both AGN and star formation are comparable in ionizing luminosity. We exclude this class for this study and note that by removing this class of potential AGN, we may be removing an important phase in the AGN-galaxy co-evolution at low redshift. The main problem with the composite class is the lack of easily accessible spectral indicators that would allow a clean separation of AGN and star formation.

We are thus left with galaxies whose emission lines are dominated by sources of ionization other than young stars. These sources may be AGN, but can also include slow-moving shocks and gas excited by post-AGB stars and horizontal branch stars \citep[see][]{2008ARA&A..46..475H, 2008MNRAS.391L..29S, 2009arXiv0912.0275S}. These non-stellar sources are empirically divided into two branches, a division that is most obvious in the \OIHa\ diagram (Figure \ref{fig:bpt}, \textit{right}). The lower branch of this population was first designated as `low-ionization narrow emission-line regions', or LINERs, by \cite{1980A&A....87..152H}. The upper branch is identified with galaxies hosting Type 2 (Seyfert) AGN.

Originally thought to be low-luminosity AGN, the nature of the LINER ionisation mechanism has since been debated, and it is becoming increasingly clear that the majority of galaxies with LINER spectra are unlikely to be low-luminosity AGN. The physical footprint of the line emitting area in nearby LINERs has been determined to be extended on kpc scales \citep[e.g.,][]{1983ApJ...268..632K, 1986AJ.....91.1062P, 2006MNRAS.366.1151S} and thus is unlikely be photoionized by a nuclear source. A detailed analysis of the distribution of this extended LINER emission in nearby \textsc{SAURON} galaxies by \cite{2009arXiv0912.0275S} demonstrates that the majority of LINERs are incompatible with an ionization source in the nucleus, even in the case of galaxies known to host (X-ray or radio-detected) AGN, but instead are powered by some component of the underlying old stellar population. \cite{2009arXiv0912.0275S} further show that this extended LINER emission is sufficiently strong to be detected in SDSS spectra of more distant $z \sim 0.05$ galaxies. \cite{2009arXiv0912.1643C} have recently similarly argued that many of the SDSS objects in the LINER box are ``retired'' galaxies powered by old stellar populations.

While it is certainly the case that some X-ray detected AGN exhibit \textit{nuclear} LINER spectra \citep[e.g.,][]{2009A&A...506.1107G}, the presence of LINER-like line ratios \textit{over the large physical footprint of the SDSS spectroscopic fiber} does not imply the presence of a low-luminosity AGN. LINERs are thus potentially highly diverse and the majority of them are not low-luminosity AGN. Even if a low-luminosity AGN were present, the \LOIII emission sampled by the spectroscopic fiber cannot disentangle the relative contributions. We therefore decided to eliminate them from our sample and to retain only the definite AGN for our analysis. 

Since galaxies exhibiting LINER spectra are different from those hosting Seyfert AGN, we verify that the inclusion of the small number of LINERs that are above the luminosity limit of \LOIII\ $\sim 10^{40}$\ergs established in the next Section does not significantly change our results, as most LINERs have substantially lower \OIII\ luminosities and Eddington ratios compared to Seyfert AGN and are thus not sites of substantial black hole growth (see Appendix \ref{sec:liners} for details). 

The \OIHa\ diagram allows for the best separation of AGN and LINERs, so we use it for those spectra where \OI\ is detected with S/N $>$ 3. In cases where \OI\ is not detected, but \SII\ is, we use the \SIIHa\ diagram (Figure \ref{fig:bpt}, \textit{center}). In both cases, we use the AGN-LINER division of \cite{2006MNRAS.372..961K}. If neither \OI\ nor \SII\ is detected, but \NII\ is, we make use of the original \NIIHa\ diagram, where we use the empirical division line defined by \cite{2007MNRAS.382.1415S}. 

Out of a parent sample of 47,675 galaxies, we thereby select a sample of 942 narrow line AGN (Table \ref{tab:sy_prop}). This corresponds to a lower limit on the AGN fraction of 2\%, as we do not include broad-line AGN, highly obscured AGN or LINERs. As we show later in this paper, the AGN fraction is highly dependent on position in the color-mass diagram and ranges from effectively zero to $\sim10\%$. 

\begin{figure*}[!ht]
\begin{center}

\includegraphics[angle=90, width=0.49\textwidth]{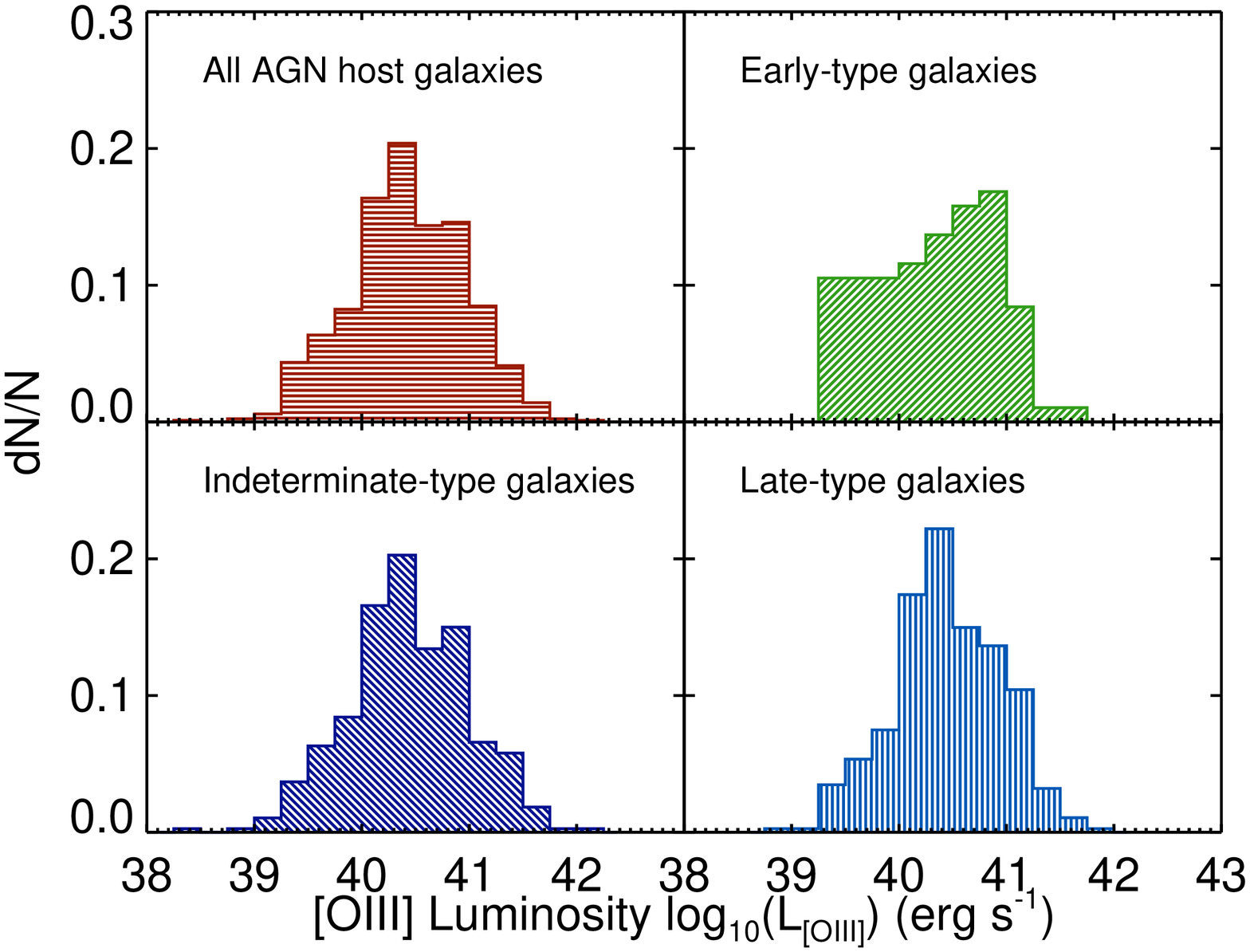}
\includegraphics[angle=90, width=0.49\textwidth]{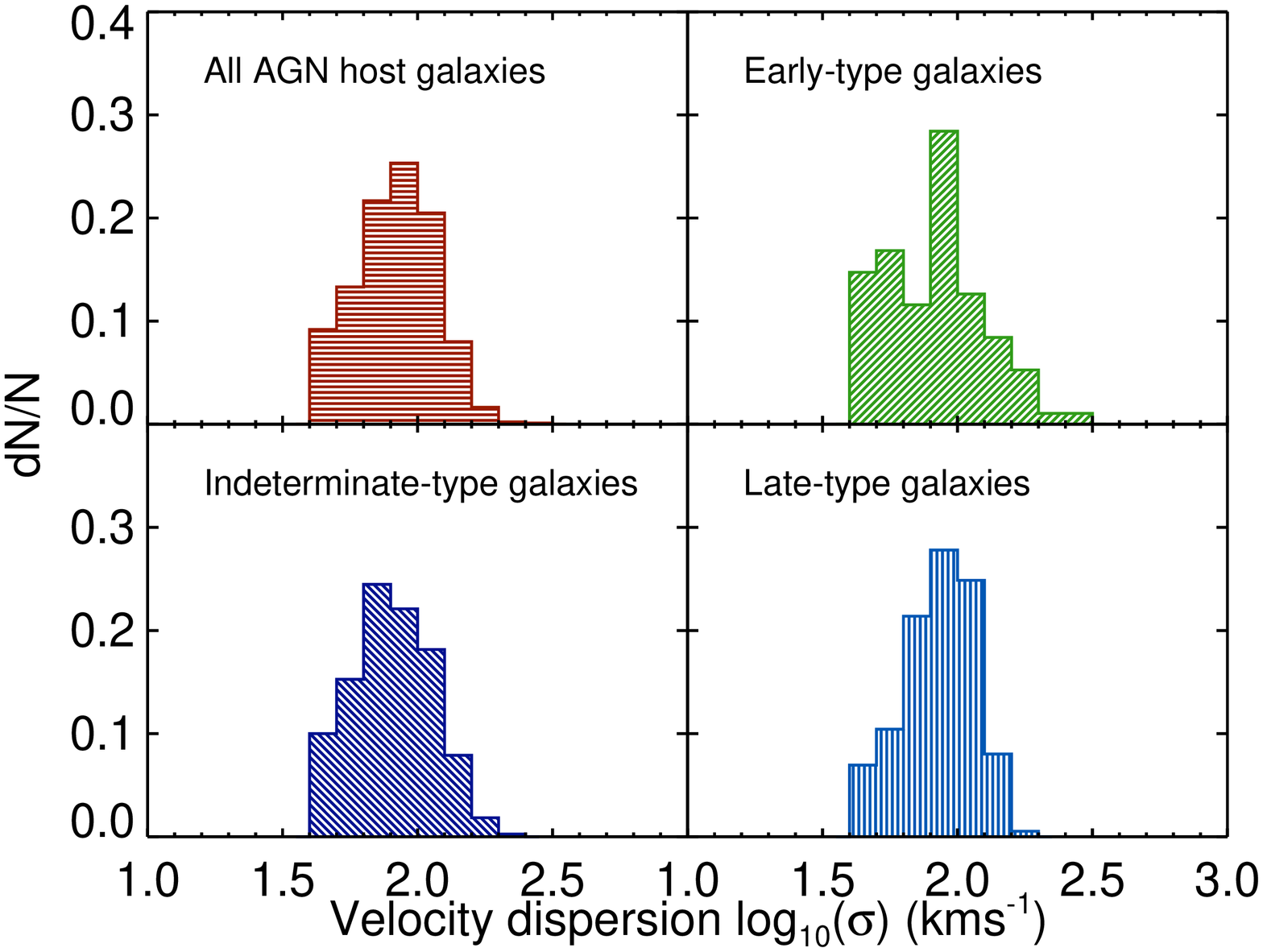}\\
\vspace{0.2in}
\includegraphics[angle=90, width=0.49\textwidth]{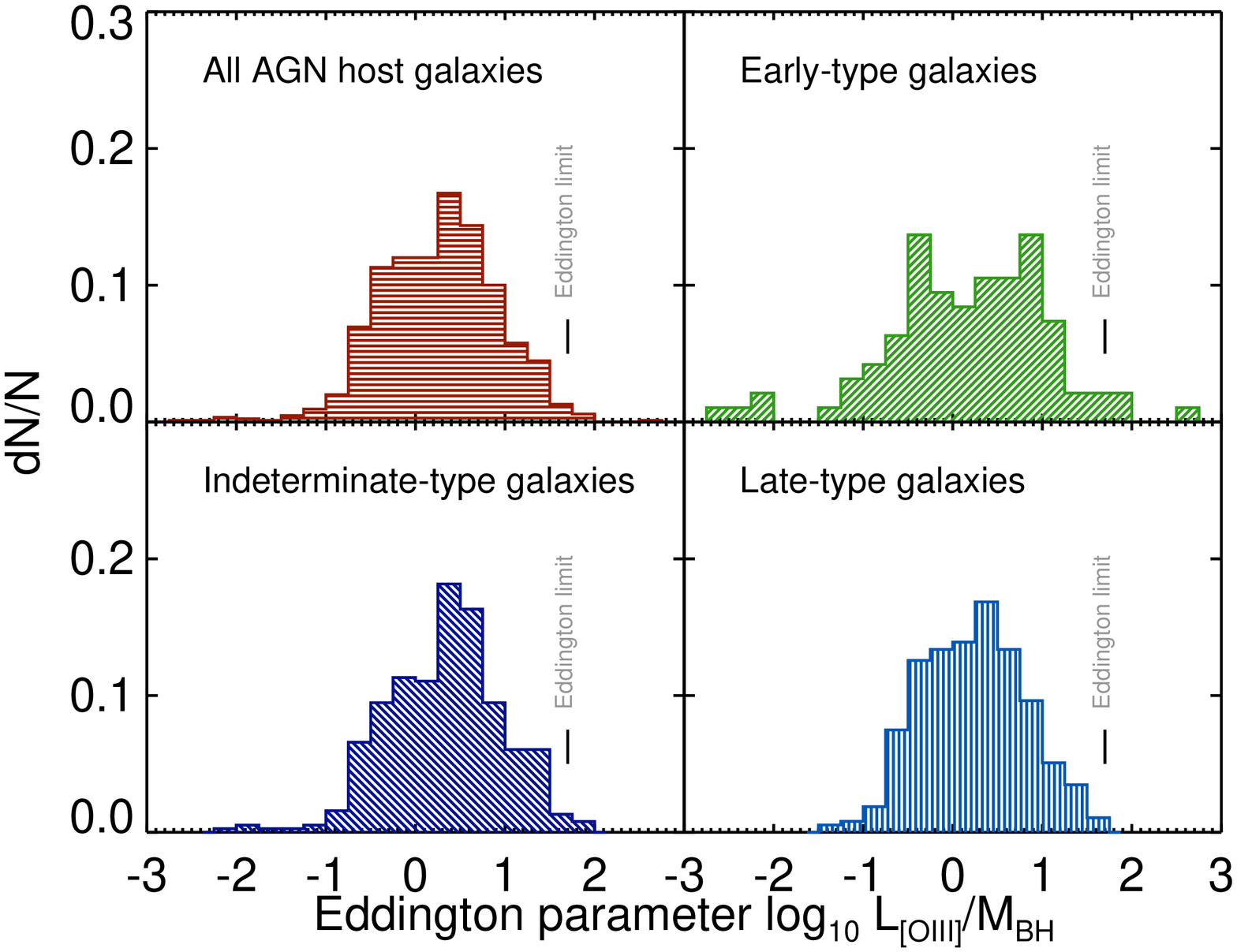}
\caption{\textit{Top left:} The distribution of \LOIII\ of our AGN sample. \textit{Top right:} The distribution of stellar velocity dispersions of our AGN sample. \textit{Bottom:} The distribution of the Eddington parameter \LOIIIMbh, combining the values from the plots on the \textit{top} row via the \Msigma\ relation. We indicate the approximate value of the Eddington limit in each panel. In each of the panels,  we show the distribution of the 
entire AGN population in the \textit{top-left}, while the remaining panels show the \LOIII\ distribution split by host morphology. \label{fig:sy_lo3}}

\end{center}
\end{figure*}

\begin{deluxetable*}{lllcccccc}[!h]
\tablecolumns{9}
\tablewidth{0pc}
\tabletypesize{\scriptsize}
\tablecaption{Typical \OIII-luminosities, Eddington ratios and black hole masses of AGN by host galaxy morphology}
\tablehead{
 \colhead{Host galaxy} & 
 \colhead{Number} & 
 \colhead{Percentage of} &
 \colhead{mean} &  
 \colhead{median} &  
 \colhead{mean} &  
 \colhead{median} &  
 \colhead{mean} &  
 \colhead{median} \\
 \colhead{morphology} & 
 \colhead{} & 
 \colhead{host galaxies} &
 \colhead{log \LOIII\ } &  
 \colhead{log \LOIII\ } &  
 \colhead{log \LOIIIs4} &  
 \colhead{log \LOIIIs4} &  
 \colhead{log \Mbh\ } &  
 \colhead{log \Mbh\ }\\
 \colhead{} & 
 \colhead{} & 
 \colhead{} &
 \colhead{(\ergs)} &  
 \colhead{(\ergs)} &  
 \colhead{} &  
 \colhead{} &  
 \colhead{(\Msun)} &  
 \colhead{(\Msun)}
}
\startdata
Late-types 				& 402	&  42.68\%	&40.40	&       40.39 &  0.295	 & 0.306 & 6.52 &        6.63\\
Indeterminate-types 	& 431	&  45.75\% 	&40.40	&       40.39 &  0.456 & 0.442 & 6.52 &        6.63\\
Early-types				& 109	&  11.57\%	& 40.27&       40.30 &  0.326 & 0.397 & 6.36 &        6.45\\
\hline
All 							& 942 	& 	100\% 		& 40.38 &      40.38 &  0.372 & 0.370 & 6.43 &        6.53\\
\enddata
\label{tab:sy_prop}
\end{deluxetable*}

\subsection{Assessing the Completeness of Emission Line Selected AGN}
\label{sec:complete}

Since we wish to study the properties of AGN host galaxies as a class, we need to understand how effective our AGN selection technique is and to what degree it is biased. Emission line selection and characterisation of AGN has been successfully employed in many works \citep[e.g.,][]{1981PASP...93....5B,1987ApJS...63..295V, 1995ApJS...98..477H, 1997ApJS..112..315H, 1997ApJ...487..568H, 1997ApJS..112..391H, 2003MNRAS.346.1055K, 2007MNRAS.382.1415S, 2008ApJ...673..715C, 2006MNRAS.372..961K}. Deep \Chandra\ studies of such emission-line selected AGN by \cite{2001ApJ...549L..51H} show that the majority of such AGN also exhibit a nuclear X-ray source. For emission line selection of AGN to be reliable and reasonably complete, two assumptions must be true:

\begin{enumerate}
\item the AGN must always excite a narrow-line region, and
\item the emission lines from the narrow-line region must never be overwhelmed by lines excited by other processes, in particular star formation.
\end{enumerate}

Many recent studies of AGN host galaxies have noted that AGN tend to concentrate in the green valley on the color-mass diagram, i.e., at intermediate optical colors, and that there are very few AGN in the blue cloud and the red sequence. This raises the possibility that AGN in star-forming blue cloud galaxies could be overwhelmed by emission from star formation. The lack of relatively luminous AGN ($L_{\rm X} \sim10^{43-45}$\ergs) in the blue cloud has been recently been established by \cite{2009ApJ...692L..19S} using hard X-ray-selected AGN, demonstrating that, at these luminosities, the absence of AGN is not a selection effect.

But could there still be a population of low-luminosity emission line AGN in the blue cloud whose emission lines are overwhelmed by star formation? For substantially lower luminosity AGN than those studied by by \cite{2009ApJ...692L..19S}, any X-ray detection in blue cloud star-forming galaxies is likely ambiguous in the 0.2--10 keV window accessed by \XMM\ and \Chandra\, as this could be due to star formation and X-ray binaries \citep[e.g.,][]{2004A&A...419..849P}. Can we do any better with emission line diagnostics?

In order to test whether emission line AGN similar to those we select on BPT diagrams might be hidden in blue cloud star-forming galaxies, we perform a simple empirical experiment. We pair a randomly selected star-forming galaxy from our galaxy sample with another randomly selected AGN. We rescale the AGN emission line fluxes to a specific \LOIII\ to simulate the effect of adding a AGN of a certain luminosity. For example, we compute the composite \NII\ line from the line of a random star-forming galaxy  and the line from a random AGN as:

\begin{equation}
[\mbox{N\,{\sc ii}}]_{\rm model} = [\mbox{N\,{\sc ii}}]_{\rm SF}  + \left( \frac{L_{\rm s}}{L_{\rm AGN}}  \right) [\mbox{N\,{\sc ii}}]_{\rm AGN}
\end{equation}

\noindent where $\rm L_{\rm s}$ is the specific \LOIII\ to which the AGN contribution to \NII\ is scaled.

We perform this pairing of star-forming and AGN host galaxies 1,000 times scaling the AGN contribution to $\rm L_{\rm s}$ $=10^{39}$, $10^{40}$ and $10^{41}$\ergs\ in \LOIII, to cover the range of star formation rates\footnote{The definition of ``star forming" depends on the detection of emission lines with sufficient signal-to-noise and does not directly correspond to a star formation rate. 95\% of star-forming galaxies have star formation rates greater than $\sim0.2$\Msunyr .} and line ratios of blue cloud star-forming galaxies and of AGN\footnote{Each pairing of an AGN with a star-forming galaxy is random, but we only have 942 AGN, so some of the AGN are used more than once.}. We show the results of this exercise in Figure \ref{fig:bpt2}, where each panel shows a \NIIHa\ diagram with the entire galaxy sample as shades. On top of each, we plot the 1,000 composite objects scaled to the three specific \LOIII\ values. This addition of an AGN component moves the position of a star-forming object off the star-forming locus, into the composite region and finally onto the AGN region. For added Sefyerts with \LOIII\ $=10^{41}$\ergs, we find that only 1.4\% remain below the \cite{2003MNRAS.346.1055K} empirical starburst line. However, for \LOIII\ $=10^{40}$\ergs and $=10^{39}$\ergs, this fraction increases to 12.8\% and 56.6\% respectively. Similarly, the fractions of AGN that move below the \cite{2001ApJ...556..121K} extreme starburst line into the composite region are 85.5\%, 54.4\% and 12.9\% respectively. 

From this experiment, we conclude that towards the higher luminosities in our sample ($\sim10^{41}$\ergs), our sample is complete even in the blue cloud, and that there is no substantial population of such blue-cloud AGN. For lower luminosities, \LOIII\ $\lesssim 10^{40}$\ergs, this is no longer true, and there may potentially be AGN in the blue cloud whose narrow-line region is overwhelmed by current star formation. If there is such a population of AGN, they cannot be detected via emission line diagnostics, just as selection in the X-rays would be challenging. However, a recent search for  for broad-line AGN hidden in star-forming galaxies by \cite{2009arXiv0904.4328M}  found only three such objects in a parent sample of over 3,200 objects, establishing a lower limit on the very low-luminosity AGN fraction in the blue cloud of $\sim0.1\%$, which suggests that the true abundance of such objects is very low, though broad lines may be similarly diluted by host galaxy light.

\subsection{Demographics of Black Hole Masses and Accretion Efficiencies}
\label{sec:sy_prop}

We diagnose the accretion state of the black hole using the \OIII\ line. Observed correlations between \LOIII\ and other indicators of AGN luminosity imply that \LOIII\ is a reasonably good indicator of the current accretion rate, though with considerable scatter \citep{2004ApJ...613..109H}. More recent work by \cite{2009ApJ...698..623D} comparing a variety of indicators of the intrinsic bolometric output of local AGN do show that \LOIII\ may not be an isotropic indicator and that Type 2 AGN -- such as those in our sample -- are intrinsically more obscured, and therefore more luminous, than previously thought, even based on hard X-ray observations (\citealt{2009ApJ...700.1878R}; cf., \citealt{2009A&A...504...73L}). Given these caveats, \LOIII\ will be used cautiously as a probe of black hole accretion in this work, and we restrict its use to statements about populations and trends rather than about individual objects. \LOIII\ is the only indicator of the current accretion rate in low-luminosity AGN available for large SDSS samples.

We infer supermassive black hole masses indirectly from the stellar velocity dispersion at the effective radius via the \Msigma\ relation \citep{2000ApJ...539L..13G, 2000ApJ...539L...9F,2002ApJ...574..740T, 2004ApJ...604L..89H} using the slope and normalisation of \cite{2002ApJ...574..740T}. While the resolution of the SDSS spectra and the templates used is approximately 60 \kms, we are able to use \textsc{GANDALF} to probe slightly below this resolution limit. Whenever we quote black hole masses, we include all objects with velocity dispersion measurements down to 40 \kms, but caution that in the range from 40 -- 50 \kms ($\log$ \Mbh $= 5.3-5.7$), the quoted black hole masses are increasingly uncertain. The median error in this lowest velocity dispersion range is $\sim 10$ \kms, corresponding to an error of $\sim 0.4$ dex in \Mbh. 

To estimate the Eddington ratio of the AGN, we adopt the measure \LOIIIMbh\ as a rough proxy \citep{2006MNRAS.372..961K}. \cite{2009MNRAS.397..135K} normalize the Eddington parameter to argue that the Eddington limit corresponds to $\log$(\LOIIIMbh\ ) = 1.7 dex. Our AGN sample contains 33 (3.5\%) objects with $\log$\LOIIIMbh\ $>$ 1.7 dex, though the large scatter in \LOIIIMbh\ makes it likely that none of our AGN are super-Eddington. \cite{2009MNRAS.397..135K} argue that the bolometric correction for the extinction-corrected \OIII\ luminosity is a factor of 300 -- 600, so the bolometric luminosities of the AGN in our sample are between 2.5 and 2.8 dex higher than the \OIII\ luminosities quoted throughout.

We summarize some of the key AGN properties, including the AGN fraction in each morphology class, in Table \ref{tab:sy_prop} and plot the distribution of the stellar velocity dispersion, \LOIII\ and \LOIIIMbh\ for both the entire AGN sample, and for each morphology class, in Figure \ref{fig:sy_lo3}. We note that the majority of the AGN in our sample ($\sim90\%$) reside in indeterminate- and late-type hosts. 

\begin{figure*}
\begin{center}

\includegraphics[angle=90, width=\textwidth]{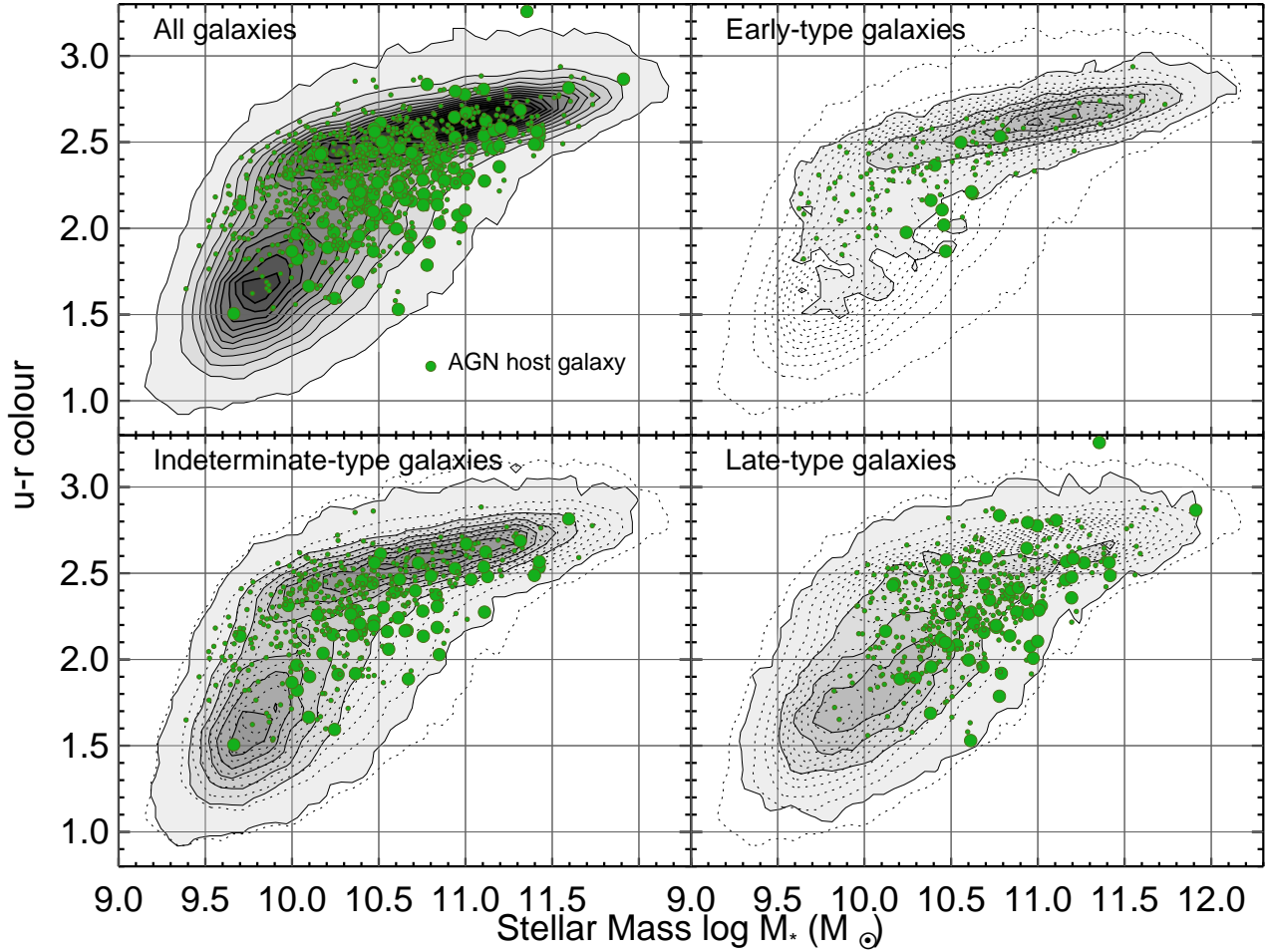}
\caption{The distribution of AGN host galaxies on the $u-r$ color-mass diagram (\textit{green dots}). \textit{Top left}: galaxies of all morphologies; \textit{top right}: early-type galaxies; \textit{bottom left}: indeterminate-type galaxies; \textit{bottom right}: late-type galaxies. The solid, shaded contours in each case show the galaxy population, on top of which we plot the individual AGN as green points. The large points are those AGN with \LOIII $> 10^{41}$\ergs. In the panels for each specific morphology class, we only plot AGN host galaxies in that specific class. Furthermore, to guide the eye, we overplot the contours of all galaxies from the \textit{top left} panel as gray dotted contours. The contour levels in all panels are linear and represent the same contour levels. From this Figure, we see that AGN host galaxies preferentially have green host galaxy colors and a range of stellar masses -- but that this range is clearly morphology-dependent. That is, AGN host galaxies are a very particular sub-set of normal galaxies, and the way in which they are separate depends on morphology.\label{fig:sy_cmass}}

\end{center}
\end{figure*}

\section{Two Distinct Modes of Black Hole Growth in Early- and Late-type AGN Host Galaxies}
\label{sec:results}

With the various observed and derived quantities assembled, we can now carry out a systematic investigation of the nature of AGN host galaxies and thus assess the role of AGN in the ongoing evolution of the low-redshift galaxy population. In Figure \ref{fig:sy_cmass}, we show the location of AGN host galaxies on a color-mass diagram of $u-r$ color vs. stellar mass as calculated in Section \ref{sec:stellarmass}. In each panel, we show the entire normal galaxy population as contours. On top of the normal galaxy distribution, we overplot the 942 AGN host galaxies as individual green points to show their absolute distribution in the color-mass diagram. The \textit{top left} panel contains the entire sample regardless of morphology. Going clockwise around the Figure, the \textit{top right} panel shows only early-type galaxies, both for the normal galaxy population and the AGN host galaxies. Below in the \textit{bottom-right}, we show only the late-types and finally in the \textit{bottom left}, the indeterminate-types that did not receive a $>80\%$ classification. In the panels split by morphology, we  overplot the contours of the entire population from the \textit{top left} panel as gray dotted contours to aid the eye in identifying the underlying blue cloud and red sequence.

The absolute distribution of AGN host galaxies on the color-mass diagram conveys the range of observed host properties, but can obscure the equally significant \textit{relative} incidence of AGN in various areas on the color-mass diagram. Previous studies have only used this absolute distribution to interpret the role of AGN in galaxy evolution -- and then only for the whole galaxy population, not for the two main morphology classes separately. 

The AGN  \textit{fraction} on the color-mass diagram (Figure \ref{fig:sy_frac}) is more enlightening than the absolute distribution because in any area on the color-mass diagram, \textit{the AGN fraction is a proxy for the duty cycle of AGN and therefore for the importance of AGN for the evolution of galaxies of specific masses, colors and morphologies}. The duty cycle is the fraction of the time that a galaxy spends as a AGN in a specific population. The AGN fraction is a proxy of this because galaxies do not move rapidly on this diagram relative to the life time of AGN phases. Typical AGN lifetimes are estimated to be on the order of $10^6 - 10^8$ years, approaching $10^9$ years only for the most massive black holes which, in any case, are not active in the local Universe \citep{2001ApJ...547...12M, 2004MNRAS.351..169M,2004cbhg.symp..169M}. The movement of galaxies on the color-mass diagram on the other hand is slow: mass-doubling timescales due to star formation are  on the order of $10^{9} - 10^{10}$ years \citep[][]{2004MNRAS.351.1151B, 2007ApJ...660L..43N} while major mergers are rare in the local Universe  \citep{2009arXiv0903.4937D}. Stellar evolution limits the pace of movement in the color direction. The most rapid color evolution possible is that of the instantaneous shutdown of star formation resulting in a movement from blue to red in $\sim1 $Gyr. Thus, the fraction of AGN in any area on the color-mass diagram is probing the fraction of time that AGN are switched on in that population, and therefore it is a measure of the duty-cycle. 

We present the AGN fraction for our sample in Figure \ref{fig:sy_frac}, which has a very similar layout as Figure \ref{fig:sy_cmass}. We do not fill the general galaxy population contours, but do leave the gray dotted contours indicating the whole galaxy population to guide the eye. We overplot the fraction of AGN host galaxies, with increasingly dark shades of green indicating a higher AGN fraction. It is immediately apparent from Figures \ref{fig:sy_cmass} and \ref{fig:sy_frac} that the absolute and relative distribution of AGN host galaxies are substantially different, and that this difference is a very strong function of morphology.

\subsection{The Importance of AGN Host Galaxy Morphology}
\label{sec:hostgals}

\begin{figure*}
\begin{center}

\includegraphics[angle=90, width=\textwidth]{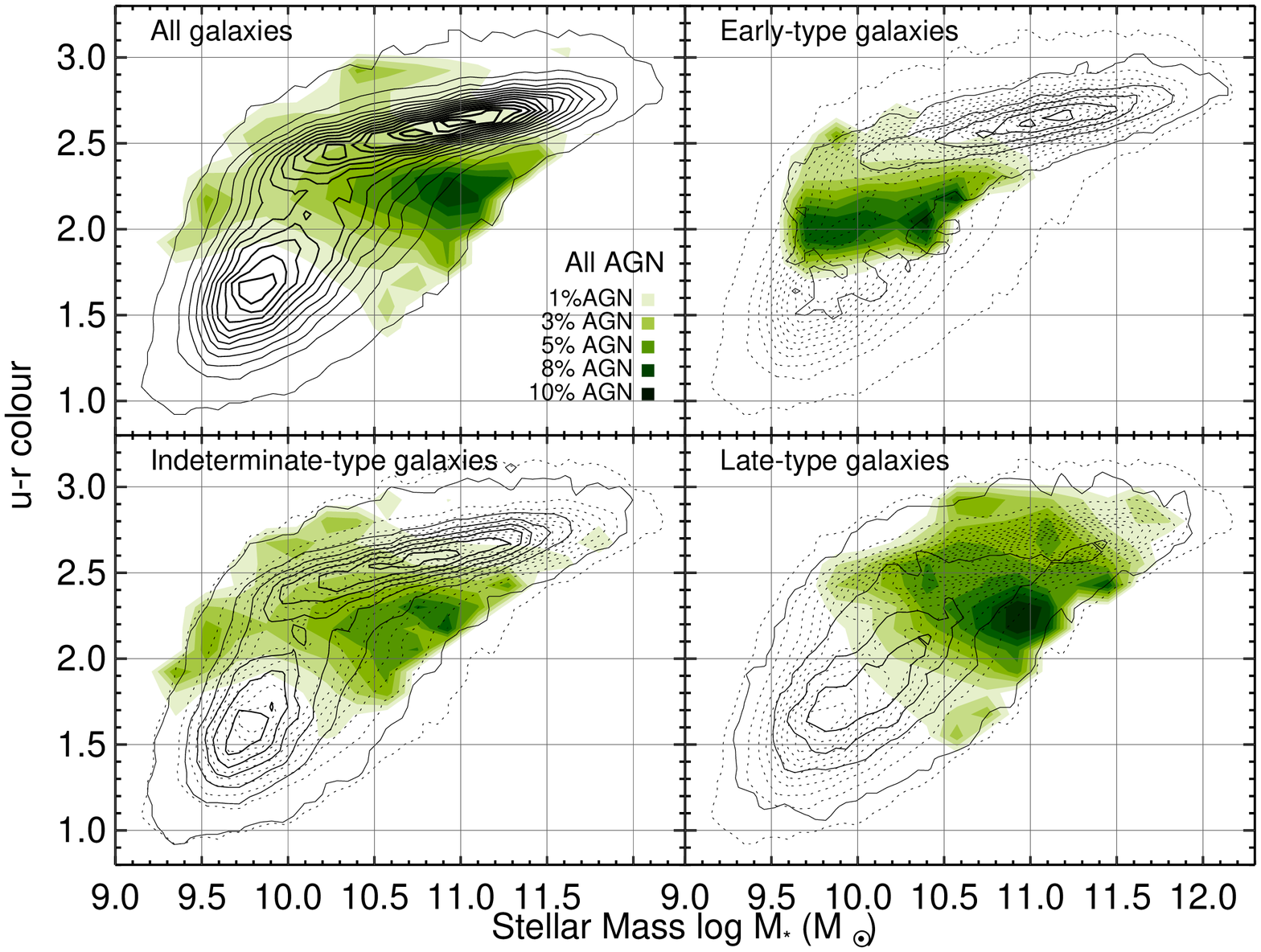}
\caption{The distribution of the \textit{fraction} of galaxies that host AGN on the color-mass diagram. The layout of this figure is similar to that in Figure \ref{fig:sy_cmass}, though for clarity, we do not shade the contours of the galaxy population. The filled contours represent the AGN fraction, which has been calculated in cells of size 0.175 dex in stellar mass and 0.125 mag in $u-r$ color. We consider only cells that contain at least 50 objects to minimize noise. The legend in the \textit{top left} panel relates the shading level to the AGN fraction. As we argue in Section \ref{sec:results}, the AGN fraction is a proxy for the AGN duty cycle. This Figure reveals in in stark terms that those galaxy populations that only very specific sub-populations of galaxies have a high AGN duty cycle: in the early-type population, it is only low-mass galaxies in the green valley that have a high AGN duty cycle. Early-type galaxies on the red sequence, regardless of mass, do not have an appreciable duty cycle. In the late-type population, it is massive galaxies that have a AGN high duty cycle. The duty cycle peaks strongly in a `sweet spot' in the green valley. The high mass of late-type AGN host galaxies is somewhat misleading, as they have similar black hole masses, and therefore bulge masses, to the early-type AGN hosts. Their higher stellar mass is therefore due to a disk, rather than a bulge (see Table \ref{tab:sy_prop}). Example images are shown in Figure \ref{fig:massive_late_type_agn}. The Milky Way galaxy most likely resides in the `sweet spot' for the AGN duty cycle in late-types. \label{fig:sy_frac}}

\end{center}
\end{figure*}

\begin{figure*}
\begin{center}

\includegraphics[angle=90, width=\textwidth]{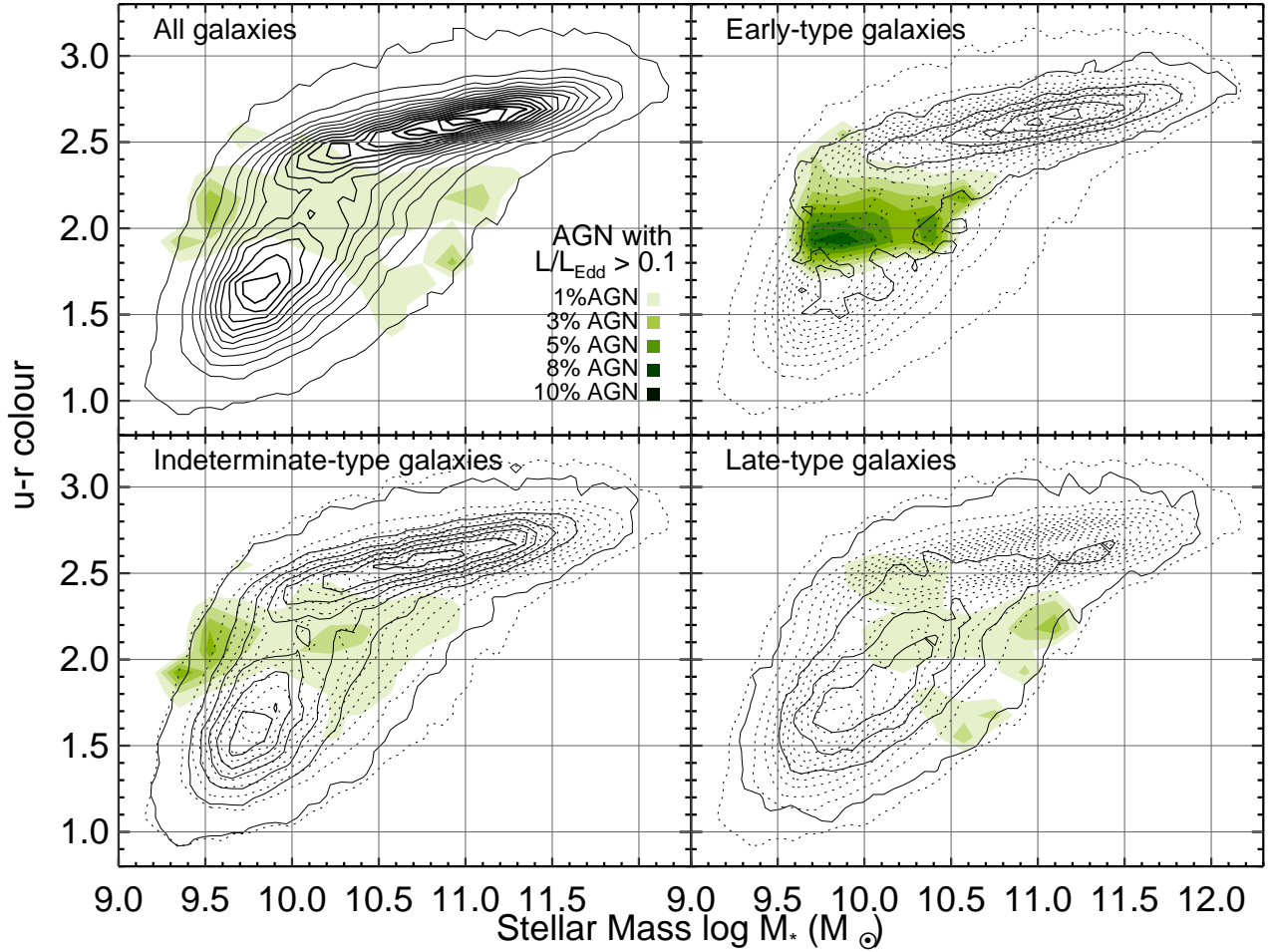}
\caption{The distribution of the fraction of AGN host galaxies on the color-mass diagram for AGNthat have an Eddington parameter \LOIIIMbh $> 0.7$ which corresponds to an Eddington ratio of $\gtrsim0.1$. The layout is identical to Figure \ref{fig:sy_frac}. This Figure shows that the duty cycle of high Eddington AGN is very low in all galaxy populations except for low mass early-types in the green valley. The prominent locus of high duty cycle in massive late-types in Figure \ref{fig:sy_frac} is therefore not dominated by high Eddington accretion.\label{fig:sy_frac2}}

\end{center}
\end{figure*}

\begin{figure*}[!ht]
\begin{center}

\includegraphics[angle=90, width=0.75\textwidth]{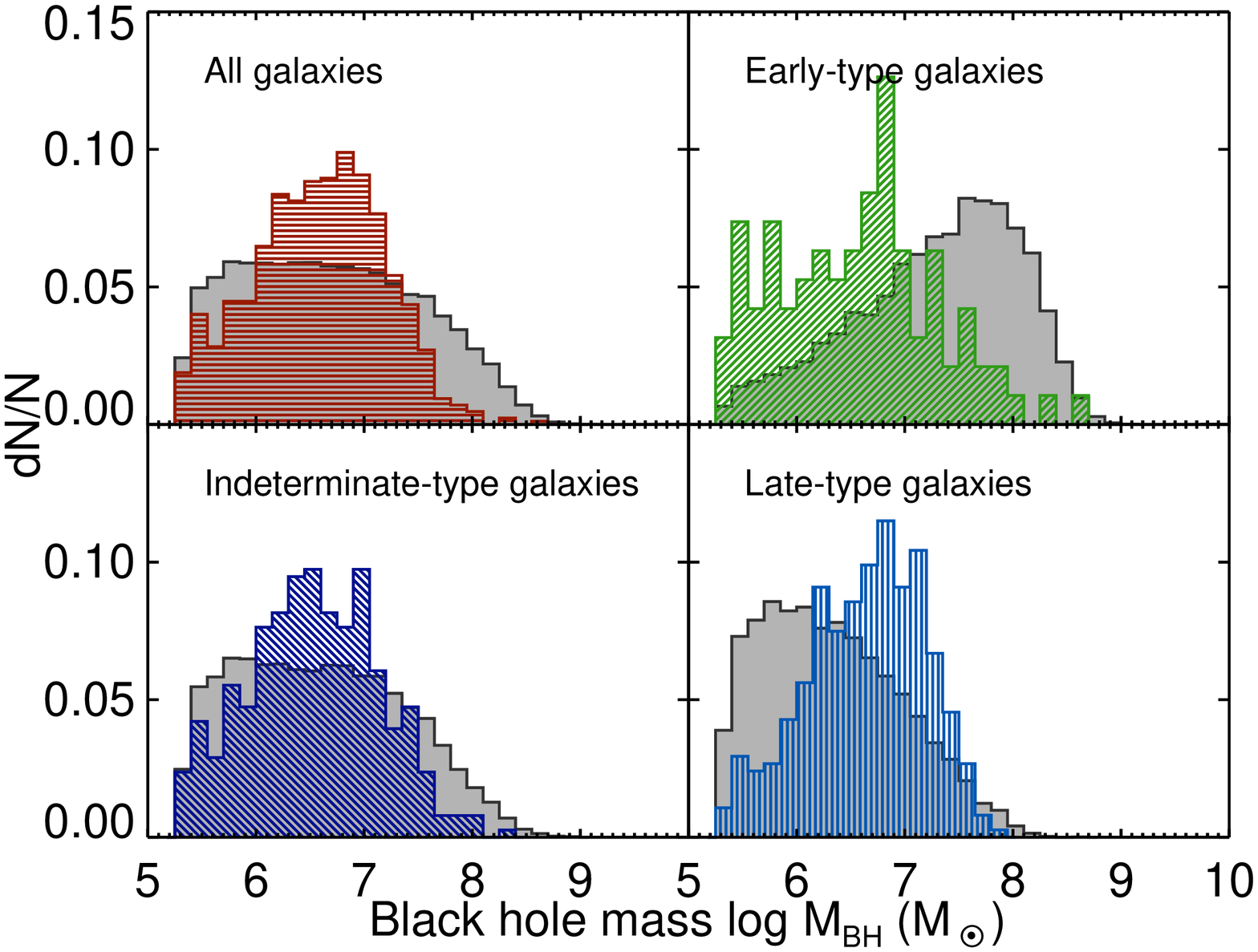}
\caption{The distribution of supermassive black hole masses \Mbh\ for both normal galaxies and AGN host galaxies as a function of morphology. In each panel, we show histogram of \Mbh\ for the normal galaxy population (\textit{gray, solid}) and the AGN only (\textit{colored, dashed}). Both here and in Figure \ref{fig:sy_fmbh}, we plot only objects where the measured velocity dispersion is greater than 40\kms (corresponding to log(\Mbh) $\sim5.3$; \citealt{2002ApJ...574..740T}). The mean and median black hole masses of both early- and late-type AGN host galaxies are very similar (see Table \ref{tab:sy_prop}), but when their distributions are compared to those of normal galaxies, large differences become apparent. Among the early-type galaxy population, it is preferentially those galaxies with lower black hole masses that are AGN, while the opposite is the case for the late-type galaxies.\label{fig:sy_mbh}}

\end{center}
\end{figure*}

 Figures \ref{fig:sy_cmass} and \ref{fig:sy_frac} suggest that the early-type and late-type host galaxies of AGN are fundamentally different populations, and that analysing the entire AGN host galaxy population as a whole obscures these differences. There is no `typical' AGN host galaxy.

We note at this point that, out of the entire sample of 942 AGN host galaxies, only 12 (1.3\%) were classified as ongoing major mergers with serious disturbances by \cite{2009arXiv0903.4937D}; the vast majority of emission-line selected AGN in the local Universe are \textit{not} hosted by galaxies undergoing a major merger.

From their distribution in Figure \ref{fig:sy_cmass}, we see that the AGN host galaxies span nearly the entire observed range of the general galaxy population (in this sample) in stellar mass. This is not the case for color; many works have already noted that AGN host galaxies preferentially reside in the green valley at intermediate host galaxy colors and that they are less common in the blue cloud and the red sequence \citep[e.g.,][]{2007MNRAS.382.1415S, 2007ApJ...660L..11N, 2007MNRAS.381..543W, 2007ApJS..173..267S, 2008ApJ...673..715C, 2007ApJS..173..342M, 2008MNRAS.385.2049G, 2008ApJ...675.1025S,2009ApJ...692L..19S,2009ApJ...693.1713T}. 

The concentration of AGN host galaxies in the green valley is strongly dependent on whether we are considering the relative or absolute AGN distribution. Figure \ref{fig:sy_cmass} shows that a substantial number of AGN lie on the red sequence. However, when we consider the AGN fraction in Figure \ref{fig:sy_frac}, this population is put into context -- compared to the total number of galaxies on the red sequence, the AGN population is lower by at least a factor of 10--20 than in the peak regions, and therefore so is the duty cycle of these AGN. Hence, the importance of AGN to galaxy evolution on the red sequence is likely very small:  we note that the population of AGN host galaxies at the massive end of the red sequence is almost entirely of indeterminate- and late-type morphology; massive early-type hosts are rare. 

The AGN fraction is strongly concentrated below the red sequence and at relatively high host stellar masses of $M_{\rm *} \sim 10^{11} $ \Msun. Furthermore, the AGN fraction rapidly drops towards lower stellar masses and is very low towards the lower mass range.

The absence of emission line AGN host galaxies on the red sequence is significant, as the emission line signature of even a very low-luminosity AGN superimposed on a passive stellar population would be detected. Interpreting the absence of AGN host galaxies from the blue cloud is more complex: As discussed in Section \ref{sec:complete}, their absence is not significant for AGN at the lower end of the \LOIII\ of AGN in our sample (see Figure \ref{fig:sy_lo3}), because the signature of the AGN narrow-line region could be overwhelmed by star formation. At higher luminosities of $\gtrsim 10^{41}$ \ergs, the absence of blue cloud AGN is highly significant, as we would have been able to detect them. At even higher luminosities of $L_{\rm X} \gtrsim 10^{43}$\ergs, \cite{2009ApJ...692L..19S} have established the complete absence of AGN in the blue cloud using hard X-ray observations by the \Swift\ satellite.

What happens when we split the AGN population by morphology? In the remaining panels of Figure \ref{fig:sy_cmass} and \ref{fig:sy_frac}, a dramatically different picture emerges wherein the early- and late-type AGN host galaxies are shown to be significantly different populations. The absolute distribution of AGN host galaxies in Figure \ref{fig:sy_cmass} already shows that the typical host stellar mass early-type AGN hosts is lower than that of late-type AGN hosts. 

Turning to the AGN fraction and therefore to where the AGN duty cycle is highest, this picture becomes even more extreme. The AGN fraction peaks at a stellar mass of $M_{\rm *} \sim 10^{11} $\Msun\ in the general population, and ranges from $10^{9.5}-10^{11.5}$\Msun. Once we split by morphology in Figure \ref{fig:sy_frac}, the difference between early- and late-type hosts becomes even more striking than for the absolute distribution: while for the late-types the AGN fraction still peaks strongly at $10^{11} $\Msun, the AGN fraction in the early-type population peaks at $M_{\rm *} \sim 10^{10} $\Msun, an order of magnitude lower than in both the late-types and and the general population. This division has not been apparent in previous works due to a lack reliable morphological classification for very large numbers of galaxies. Because the early-type AGN make up only $\sim 11\%$ of all AGN (see Table \ref{tab:sy_prop}), their properties are overwhelmed by the late-types in a general AGN host galaxies sample.

The most striking aspect of the AGN fraction in Figure \ref{fig:sy_frac} is that in the early-type population, the AGN duty cycle is highest in an extreme population of early-types that do not reside on the red sequence, but rather below it at the low-mass end and above the bulk of the blue cloud, i.e., in the green valley. Similarly, the AGN duty cycle for late-type galaxies is highest in objects that are more massive and redder than the typical late-type population. These red late-types will include a substantial fraction of inclined spirals with intrinsically blue colours, but dimmed and reddened by dust. If the detected AGN fraction is independent of host galaxy inclination, which is expected if the bulk of the nuclear obscuration comes from the central regions orientated randomly with respect to the galaxy, then these contaminants to the red part of the late-type colour-mass diagram from the part with much lower AGN fractions will make this conclusion even stronger.

The indeterminate-type galaxies exhibit an AGN fraction distribution similar to the late-types, though their AGN fraction does not extend into the higher mass red sequence. This may be explained by the fact that the incidence of indeterminate-type galaxies in the red sequence is much higher than that of late-types. We therefore tentatively conclude that the indeterminate-type AGN, which are 45\% of the AGN population after all, appear to behave more similar to the late-types than the early-types. 

When we make a cut in Eddington parameter of \LOIIIMbh $> 0.7$, corresponding to an Eddington ratio of \Ledd\ $\sim 0.1$, an even more extreme picture occurs. By making a cut in Eddington ratio, we avoid any bias towards more massive black holes which radiate at higher luminosities at fixed \Ledd\ than their low-mass counterparts. Figure \ref{fig:sy_frac2} is identical to Figure \ref{fig:sy_frac} except for the cut in \Ledd. This reveals that the duty cycle of high Eddington AGN is significant in only one population of low-redshift galaxies: low-mass early-type galaxies in the green valley. 

We conclude that the early- and late-type host galaxies of AGN are fundamentally different populations: \textit{for late-type hosts, AGN preferentially reside in \underline{massive ($10^{11} $\Msun}) galaxies either with green colors or on the red sequence. In contrast, early-type AGN tend to have \underline{moderate mass ($10^{10}$\Msun)} and strongly cluster in the green valley in between the blue cloud and the red sequence. We further showed that the duty cycle for high Eddington AGN is high only in green valley early-type galaxies.}

The AGN fraction on the color-mass diagram is a new and powerful tool to probe the duty cycle of AGN. Under the assumption that a high duty cycle AGN phase is more likely to affect the host galaxy evolution, the duty cycle is in turn a measure of the importance of AGN for galaxy formation for specific galaxy populations.

\subsection{Black Hole Demographics and Eddington Ratios in Early- and Late-type AGN Host Galaxies}

\begin{figure*}
\begin{center}

\includegraphics[angle=90, width=0.75\textwidth]{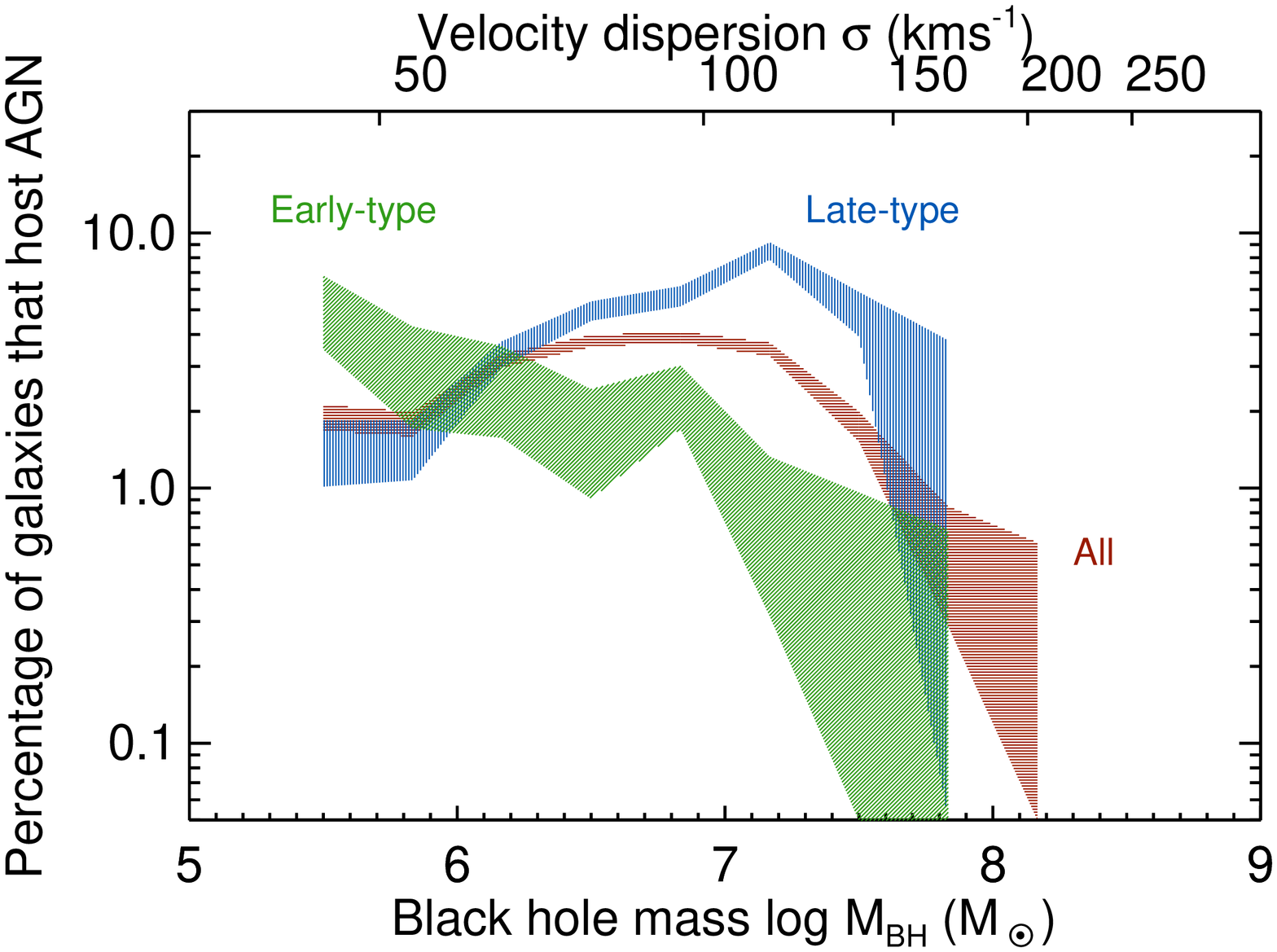}
\caption{The fraction of supermassive black holes that are growing in the local Universe. We plot the black hole mass \Mbh\ versus the fraction of galaxies that host an AGN. The errors on the fraction are Poisson errors. We plot the AGN fraction for the entire sample as well as for galaxies of specific morphology classes: (\textit{red}) all; (\textit{green}) early-types; (\textit{blue}) late-types. This plot elaborates on the trend seen in Figure \ref{fig:sy_mbh} and shows the relative incidence of AGN in galaxies of early- and late-type morphology. The AGN fraction -- and therefore the duty cycle -- in early-type galaxies is heavily skewed toward low black hole masses. In late-type galaxies, the AGN fraction is skewed towards higher black hole masses, potentially with a drop at the highest masses.\label{fig:sy_fmbh}}

\end{center}
\end{figure*}

We have shown that the properties of early- and late-type AGN host galaxies are different. This in turn strongly suggests that the triggering and fuelling of the AGN, as well as their role in the host evolution, are different. We therefore continue our discussion with the black hole masses and accretion efficiencies of black holes in early- and late-type AGN host galaxies.

Note that the characteristic black hole mass of the AGN is \Mbh\ $\sim10^{6.5}$\Msun\ and that this value is approximately independent of the host galaxy morphology (Table \ref{tab:sy_prop}).  The active black holes in late-types hosts are typically more massive than those in early-types by only $\sim$0.2 dex, or less than a factor of 2. Combined with the fact that the typical accretion rate estimated by \LOIII\ is the same in early- and late-types, we find that the Eddington ratios as inferred via the Eddington parameter \LOIIIMbh\ are similar, with the AGN in early-types having Eddington ratios that are only about a factor 1.5 lower than those in late-types. Thus, the typical black hole mass growing today in early- and late-type galaxies is the same, with the same range of Eddington ratios. In addition, the host galaxy colors of the AGN host galaxies place them preferentially in the green valley, adding to an apparent picture where the conditions for black hole growth are approximately independent of host morphology. 

However, as we have shown in the previous section, the stellar masses of early- and late-type host galaxies are very different, so taking only population averages into consideration can obscure crucial variations between populations. Similarly, it is not just relevant which black holes are currently accreting, but also to what degree these growing black holes are different from those in their quiescent parent population, and crucially, how this difference depends on host morphology. 

In Figure \ref{fig:sy_mbh}, we plot the distribution of inferred supermassive black hole masses for both the AGN (\textit{striped}) and the normal galaxy population (\textit{soild}), both for the entire sample, and split by host morphology. These histograms highlight that taking the mean, and ignoring the distribution of black hole masses in the parent population masks intrinsically very different distributions of black hole growth. Even though the median black hole masses of active early- and late-type galaxies are comparable (Table \ref{tab:sy_prop}), a Kolmogorov-Smirnov test shows that they are drawn from different parent distributions at the 99.7\% significance level.

The distribution of active black holes in late-type hosts skewed towards higher black hole mass, peaks at \Mbh\ $\sim10^{7}$\Msun\ and cuts off above \Mbh\ $\sim 10^{8}$\Msun. This contrasts sharply with the distribution of black hole masses in all late-types galaxies, which decreases with increasing black hole mass. Plotting the \textit{fraction} of active supermassive black holes in late-type galaxies in Figure \ref{fig:sy_fmbh}, we find a strongly increasing fraction from $\sim 1\%$ at $10^{5.5}$\Msun to $\sim 8\%$ at $10^{7.25}$\Msun. At greater black hole masses, the fraction may begin to decline again, until we run out of number statistics.

The black holes of early-type galaxies show a radically different picture (Figure \ref{fig:sy_mbh}).  The mass distribution of active black holes in early-type galaxies is skewed towards the lowest inferred masses in our sample. Plotting the active fraction as for the late-types in Figure \ref{fig:sy_fmbh}, we find a very different behaviour. The fraction of active black holes in early-type galaxies is \textit{decreasing} with increasing black hole mass, but we run out of galaxies to provide adequate number statistics.

These two very different trends in early- and late-type hosts are unlikely to be selection effects. Since we are able to detect AGN in low-mass early-types down to host galaxies with stellar masses of $10^{9.5}$\Msun; we do not see AGN in late-type hosts where the lowest-mass hosts are $\sim10^{9.5}$\Msun. In order to substantially change the trend of increasing AGN fraction with increasing black hole mass, we would have to have missed a very large number of AGN in low mass blue cloud late-type galaxies, more than ten times the number in our sample. To change the distribution to one skewed to low masses similar to the distribution of the early-types, an increase by a factor of $\sim$1000 would be required (Figure \ref{fig:sy_frac}). At luminosities above the $\sim10^{40}$\ergs\ completeness limit in Section \ref{sec:complete}, we are not missing any AGN in star-forming galaxies that might change this trend. Since the majority of the AGN contributing to the trends in Figure \ref{fig:sy_fmbh} are above this limit (see Figure \ref{fig:sy_lo3}), only lower-luminosity AGN might change these trends  (see also \ref{sec:liners}). Currently, there are no indications that such a large population of low-luminosity AGN exists (factor of $\sim$1000 greater in number than observed), but our results are an additional motivation to search for them.

We restrict our AGN sample by Eddington ratio as we did in Section \ref{sec:hostgals} by limiting the AGN sample to those AGN with \LOIIIMbh $> 0.7$ (which corresponds to \Ledd\ $\sim0.1$) in order to test whether the trend we see in Figure \ref{fig:sy_fmbh} is due to a bias against low-mass black holes at a fixed Eddington ratio. Such a bias is unlikely to be the cause of the different trends we see for early- and late-type hosts as it would imply vastly different distributions of Eddington ratios. Imposing this cut in Eddingto ratio to Figure \ref{fig:sy_fmbh} yields qualitatively the same trend, though the increase in the AGN fraction for the late-type hosts is less steep.

\textit{In summary, dividing the local AGN host galaxy population by morphology reveals that that there are two very different modes of black hole growth: in early-types, it is preferentially the \underline{least massive} black holes that are actively accreting material in an AGN phase. In late-type galaxies, it is preferentially the \underline{most massive} black holes that are active, with a potential decline at the very highest masses.} Since the early-type AGN are only 11\% of the population, this effect can easily be masked by studies of the AGN population as a whole. 

\section{A New View of the Co-evolution of Galaxies and their Supermassive Black Holes}
\label{sec:discussion}

Despite the differences discussed in the previous section, the velocity dispersion -- and therefore the bulge mass and black hole mass -- of AGN host galaxies is independent of morphology (Table \ref{tab:sy_prop} and Figure \ref{fig:sy_lo3}). Another common feature is the green host galaxy colors. This raises the possibility that these conditions are most favorable to fuelling a black hole, at least in low-redshift galaxies. The late-type AGN hosts are an order of magnitude more massive $\sim10^{11}$\Msun\ simply because of the presence of a massive disks. The bulges of late-types that host AGN appear to be very similar to the low-mass, $\sim10^{10}$\Msun, early-type galaxies that host AGN.

However, as we have shown, the early- and late-type host galaxies of AGN are fundamentally different populations both in terms of \textit{which galaxies} host AGN, and also \textit{which black holes} are growing. Only by comparing the AGN hosts to normal galaxies not just of the same mass and color, but also morphology, does it become clear that the two populations differ from each other in fundamental ways. 

What are the consequences of our findings for the role of AGN in galaxy evolution in the local Universe? In this section, we outline a new view of the impact of AGN on galaxy evolution and make the case that AGN perform very different roles in early- and late-type host galaxies.

\subsection{The Role of AGN in the Evolution of Early-type Galaxies}

\begin{figure*}[!ht]
\begin{center}

\includegraphics[angle=90, width=\textwidth]{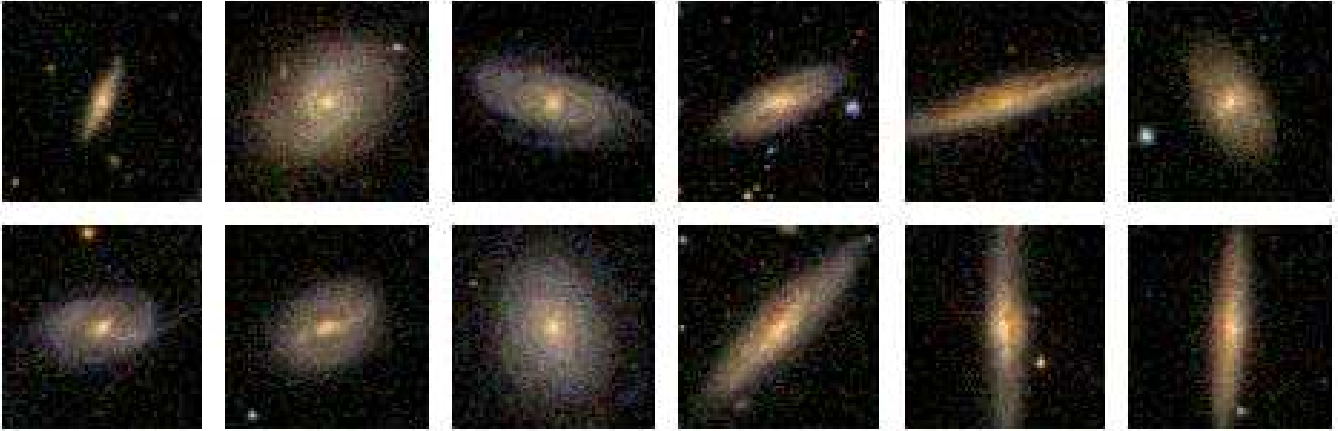}
\caption{Example SDSS $gri$ images 80\arcsec\ $\times$ 80\arcsec\ in size of late-type AGN host galaxies with stellar masses above $10^{11}$\Msun. These images illustrate that massive late-type AGN hosts are not massive bulges with a small disk (S0 or Sa), but rather have small bulges and massive disks.\label{fig:massive_late_type_agn}}

\end{center}
\end{figure*}

As discussed in Section \ref{sec:results}, while there are a number of AGN in early-type galaxies on the red sequence, the fraction of early-type galaxies on the red sequence that host AGN is very small, \textit{i.e.}, the duty cycle of AGN in this population is small as well when compared to the peak duty cycle in the green valley.

We see from Figure \ref{fig:sy_frac} that the duty cycle of AGN in early-type host galaxies is peaked in a well-defined band at intermediate green colors precisely in the green valley above the blue cloud and below the low mass end of the red sequence. If the early-type AGN hosts in the green valley are post-starburst objects, then there clearly is an ample supply of star-forming progenitor galaxies at the same mass in the blue cloud.These progenitors would evolve to the green valley if star formation is suppressed, and will continue to fade onto the low-mass end of the red sequence. Such progenitors may already have early-type morphology \citep[e.g.,][]{2009MNRAS.396..818S}, or they may first transform their morphology in a major merger, as simulations seem to show \citep[e.g.,][]{1972ApJ...178..623T,1992ApJ...393..484B,1996ApJ...471..115B}. Such a major merger may also fuel the black hole, giving rise to the AGN phase observed at a later stage \citep{1988ApJ...325...74S, 2006ApJS..163....1H}. Simulations by \cite{2009MNRAS.394.1713K} show that minor mergers with a gas-rich dwarf are not sufficient to move a red early-type progenitor to the optical blue cloud (though they may reach UV-blue colors) and so are not a good candidate for the trigger mechanism.

Of course, it does not follow from the intermediate optical colors of green valley galaxies that they must be in transition from blue to red. There are multiple, very different star formation histories that can result in green colors. A slightly enhanced dust screen covering an actively star forming galaxy would suffice. More complex scenarios are also imaginable. That early-type AGN hosts galaxies genuinely are a transition population was shown by \cite{2007MNRAS.382.1415S}; however, this does \textit{not} imply that the AGN phase in the green valley is necessarily the \textit{cause} for the shutdown of star formation and the transition from blue to red. Rather, the absence of active black hole accretion in blue cloud objects, established by \cite{2009ApJ...692L..19S}, implies that there must be a significant time delay on the order of at least 100 Myr, more likely more like 500 Myr, between whichever process suppresses star formation and the onset of substantial AGN radiation. Furthermore, \cite{2009ApJ...690.1672S} showed that the extremely rapid destruction of the molecular gas reservoir that fuels star formation coincides with a very low-level of accretion preceding the AGN phase in so-called composite objects\footnote{Composite objects lie on the BPT diagram in between the locus of star forming galaxies and the non-stellar (AGN \& LINER) locus and are usually interpreted as a combination of star formation and very low-level AGN activity.} and that this very low-luminosity phase -- perhaps accompanied by a radiatively inefficient kinetic outflow \citep[e.g.,][]{1994ApJ...428L..13N} --  may be responsible for the shutdown of star formation by removing the molecular gas reservoir.

Regardless of whether any AGN phase is responsible for the shutdown of star formation, the duty cycle of AGN in the present day early-type galaxy population is highest in low-mass green valley objects in which star formation has already been shut down and which are in the process of moving to the red sequence. The duty cycle of AGN in massive, red sequence early-types on the other hand is very low. The majority of black hole growth in early-type galaxies at low redshift thus appears to be a post-starburst phenomenon, rather than part of a coeval starburst. Figure \ref{fig:sy_frac2} furthermore illustrates that it is \textit{only} this population of low redshift galaxies that has a substantial duty cycle of high Eddington accretion.

\subsection{The Role of AGN in the Evolution of Late-type Galaxies}

The late-type AGN host galaxies are genuine late-type spirals, not S0/Sa galaxies that consist of a large bulge and a small disk, even at the highest masses, where the duty cycle is highest. We show example images of massive late-type AGN hosts in Figure \ref{fig:massive_late_type_agn}. While the host galaxies are quite massive, their bulges are not -- this is apparent from their bulge velocity dispersions, and therefore their modest black hole masses, which are comparable to those of active early-type galaxies; they therefore must have substantial disks which make up the remaining stellar mass (Figure \ref{fig:sy_mbh} and Table \ref{tab:sy_prop}). To put it another way, the typical black hole masses, bulge masses and Eddington ratios of early- and late-type AGN hosts are similar, but the late-type hosts are substantially more massive due to the presence of a massive disk.

At the same time, the distribution of Eddington parameters (Figure \ref{fig:sy_lo3} and Table \ref{tab:sy_prop}) does not provide a comprehensive picture: comparing Figure \ref{fig:sy_frac} shows that the duty cycle of AGN -- as we are able to detect them via emission line diagnostics -- is high in high-mass, green late-types. However, Figure \ref{fig:sy_frac2} shows that the duty cycle of high Eddington-rate accretion is low in those late-types, especially when compared to the green valley early-types.

Given the distinct location of late-type AGN host galaxies on the color-mass diagram, what role do the AGN play in the evolution of late-type galaxies? We argued previously that the early-type AGN represent the true green valley transition population. The majority of AGN hosted by late-type galaxies do \textit{not} reside between the blue cloud and the low-mass end of the green valley and therefore cannot be part of the same blue-green-red transition. A minority of the late-type AGN hosts do however have sufficiently low masses to plausibly originate from the blue cloud. The absence of a substantial duty cycle for high Eddington accretion in these objects further supports this, as the AGN of the early-type hosts do have such a high duty cycle. There is also an absence of AGN at the very highest masses $\gtrsim10^{11.5}$\Msun\ which may indicate a feedback process that has shut down accretion in those objects, or this may be due to insufficient number statistics.

We consider a number of scenarios that may account for the presence of AGN in late-types; any such scenario must account for three observational characteristics: (i) characteristic host mass of $\sim10^{11}$\Msun, (ii) late-type host galaxy morphology, and (iii) intermediate/green host galaxy color.

\subsubsection{Major Mergers of two Disk Galaxies}
The first possible scenario for the origin of the late-type AGN hosts is the major merger of two disk galaxies that results in the fuelling of the black hole \citep[e.g.,][]{2005ApJ...622L...9S, 2005ApJ...630..705H, 2006ApJ...645..986R}. We know from the visual inspection of merging galaxies in the SDSS Universe that disk-disk mergers at $\sim10^{11}$\Msun\ do occur \citep{2009MNRAS.tmp.1770D, 2009arXiv0903.4937D}. However, there is a lack of \textit{blue} progenitor galaxies at $\sim10^{11}$\Msun\ that might evolve toward the red sequence as star formation is suppressed (see Figure \ref{fig:sy_frac}). Both factors thus make it unlikely that late-type AGN hosts are the product of such a merger and so we conclude that this scenario does not account for criterion (i).

Also, a major merger would likely have destroyed the disk that is clearly visible in the SDSS images (Figure \ref{fig:massive_late_type_agn}). Simulations show that mergers involving particularly gas-rich progenitors can result in a disk-dominated remnant \citep{2005ApJ...622L...9S,2006ApJ...645..986R, 2009ApJ...691.1168H} , but only for particular orientation of the angular momentum vectors. Major mergers are already rare, and this requirement further reduces that probability by a large factor. Moreover, such extremely gas-rich, massive progenitor galaxies do not exist in the low-redshift Universe \citep[e.g.,][]{2004ApJ...611L..89K} and so major mergers are an unlikely channel for the triggering of AGN in low redshift late-type galaxies.

\subsubsection{Mixed-mergers of a Disk and a Spheroid Galaxy}
One possible scenario for the origin of late-type AGN is a mixed merger between a disk- and a massive spheroid-dominated system, which is predicted to be very common by some models of galaxy formation \citep[e.g.,][]{2003ApJ...597L.117K, 2006MNRAS.370..902K}. The mixed merger scenario alleviates one of the problems with major mergers, as there is an ample supply of massive early-type, red progenitors available on the red sequence. It also accounts to some extent for the green host galaxy colors (iii); while the red sequence progenitor does not have a ready supply of cold gas available for star formation, the late-type progenitor does. 

However, this scenario also faces some problems: why does the mixed merger result in green, rather than blue optical colors? One explanation might be that the late-type progenitor originates from the blue cloud proper and is substantially less massive ($\sim10^{10}$\Msun), bringing in only a small amount of cold gas that can fuel only a minor starburst. Simulations of minor mergers between passive early-type galaxies and an ensemble of small gas-rich galaxies by \cite{2009MNRAS.394.1713K} show that such events can produce the intermediate $u-r$ colors exhibited by the late-type AGN hosts. 

While a major mixed merger likely also destroys the disk in the late-type progenitor, a minor merger might not, thus satisfying criterion (ii). However, another problem is that the bulges of the late-type AGN hosts are comparable in velocity dispersion to the significantly less massive early-type AGN hosts and therefore are predominantly disk-dominated systems; a mixed merger of this type seems unlikely to build a massive disk (similar to that of the Milky Way; see Section \ref{sec:mw}) around a small bulge. Thus a contradiction arises: in order to account for the green colors, the early-type must be more massive than the late-type, but in order to ensure the survival of the disk, the late-type must be more massive than the early-type.

Another possibility is that the AGN is regulating the strength of the starburst and preventing it from moving the host galaxy all the way to blue cloud colors, which is commensurate with the idea that AGN are more common in late-type galaxies with high black hole masses. 

\subsubsection{Concluding Remarks concerning Late-type AGN Host Galaxies}
All the scenarios outlined here are problematic and none of them account for all the observed properties of late-type AGN host galaxies. Perhaps an AGN phase in late-type galaxies does not significantly affect the evolution of the host galaxy, and all we see is secular evolution. A slight increase in the amount of gas in the bulge increases the odds for stochastic accretion of gas, perhaps driven by a bar \citep{1993ApJ...409...91H, 2004ApJ...612L..17W, 2006RMxAC..26..131C} or more generally by disk-driven evolution \cite{2004ARA&A..42..603K}. Other possibilities include mass loss from evolved stars \citep[e.g.,][]{2007ApJ...671.1388D,2008ARA&A..46..475H}, the infall of a minor dwarf galaxy or the tidal disruption of a star by the black hole \citep[e.g.,][]{2006ApJ...653L..25G}

As we have discussed, it is unlikely that the late-type AGN hosts are involved in the same transition from the blue cloud to the red sequence that the early-types appear to be involved in. The role of AGN in the evolution of late-type galaxies must therefore be a different one from the early-types. A major difference between the early- and late-type AGN hosts that highlights the different nature of black hole accretion is that the late-types do not have any sub-population that has a high duty cycle of high Eddington accretion (Figure \ref{fig:sy_frac2}). Since there are no major ongoing starbursts in late-type AGN hosts and the duty cycle for high Eddington accretion is low, perhaps all back hole growth has already taken place at high redshift and all we are seeing is a minor re-activation due to one of the processes mentioned above.

Given that none of the scenario we outlined are satisfactory explanations, the co-evolution of late-type galaxies and their black holes -- unlike in the case of early-types -- remains an outstanding issue.


\subsection{Parallels between Black Hole Growth in the Present-day and Early Universe}
The only part of the early-type galaxy population that is experiencing a high duty cycle of black hole growth at high efficiencies are low-mass post-starburst early-type galaxies transitioning from the blue cloud to the low-mass end of the red sequence. This population is continuing the build-up of the red sequence, so it is natural to speculate that the process is a downsized version of the process that led to the formation and quenching of more massive early-type galaxies at high redshift which are known to be older and to have formed on shorter time scales. 

The late-type AGN are not associated with the same transition from blue cloud to the red sequence and their role in the evolution of their hosts is less clear. The majority of the black hole growth in the local Universe is associated with late-type hosts (at least $\sim40\%$, and almost 90\%, if we include all indeterminate-types), and only $\sim$11\% is associated with early-types. Thus, the mode of black hole growth at work in late-type galaxies is significantly more common in the local Universe than the one at work in early-types, and so most black hole growth is not associated with the blue--red transition seen in early-types. 

Recent observations of X-ray-selected AGN host galaxies in deep fields by \cite{2009MNRAS.397..623G} suggest that the fraction of AGN host galaxies with late-type morphology is only $\sim30\%$ at $z \sim1$, indicating a decrease of the late-type fraction with increasing redshift, though the AGN population studied by \cite{2009MNRAS.397..623G} is somewhat more luminous that in our sample. This decrease may suggest that the appearance of the mode of black hole growth at work in late-type galaxies is relatively recent, perhaps becoming apparent only around $z \sim1$ when the assembly of the Hubble sequence begins to finalize. Another possibility is that this mode has always been active at a moderate level, but that the AGN host galaxy population at high redshift is dominated by the downsizing mode at work in the early-type population. The two modes at work in early- and late-type galaxies, respectively, change in importance with cosmic time, the late-type mode may account for the the observed peak at $z\sim0.7$ in the number density of moderate-luminosity X-ray selected AGN at $z \sim 0.7$, while the mode at work in late-types is responsible for the high redshift peak of luminous AGN \citep{2001ApJ...551..624G, 2001AJ....122.2156A, 2003ApJ...598..886U, 2005A&A...430..811G}. As star formation and black hole growth moves to less massive early-type galaxies \citep[e.g.,][]{2005ApJ...621..673T, 2009arXiv0912.0259T}, the black hole growth mode in early-type galaxies also becomes less luminous and thus by $\lesssim 1$ becomes overwhelmed by the late-type mode.

Perhaps the local Universe does not tell us much about the AGN-host galaxy connection because black hole growth rates and and star formation rates have declined so substantially since $z\sim1$ except in a few rare systems more reminiscent of activity at high redshift (e.g. NGC 6240)? Further work on the morphology of AGN host galaxies at high redshift, in particular detailed restframe optical morphological classification, will be required to investigate these suggestion. Near-IR imaging data from the new Wide Field Camera 3 on board \textit{Hubble} could be used to reproduce this work at $z\sim2$. 

\begin{deluxetable*}{llcclllcccc}
\tablecolumns{9}
\tablewidth{0pc}
\tabletypesize{\scriptsize}
\tablecaption{A catalogue of AGN host galaxies}
\tablehead{
 \colhead{ID} & 
 \colhead{SDSS Object} & 
 \colhead{RA} &
 \colhead{Dec} &  
 \colhead{GZ} &  
 \colhead{Redshift} &  
 \colhead{$M_{\rm stellar}$} &
 \colhead{$\sigma$} & 
 \colhead{$u-r$} & 
 \colhead{log\LOIII} &
 \colhead{log\LOIIIs4} \\
 \colhead{} &
 \colhead{ID} &
 \colhead{(J2000)} &
 \colhead{(J2000)} &
 \colhead{class$^1$} &
 \colhead{} &
 \colhead{\Msun} &
 \colhead{\kms} &
 \colhead{colour} &
 \colhead{\ergs} &
 \colhead{} 
}
\startdata
         1   &   587722983902544102   &   14 17 04.4 &+00 30 28.8   &   i   &     0.025730   &        10.01   &         69.9 $\pm$        4.91   &         2.51   &        39.53   &       -0.181   \\
         2   &   587724197207277638   &   01 31 01.1 &+13 03 15.7   &   e   &     0.020734   &        10.80   &        115.5 $\pm$        3.23   &         2.60   &        39.33   &       -1.308   \\
         3   &   587722984428929165   &   12 40 53.6 &+01 00 30.1   &   e   &     0.022932   &        10.76   &         73.7 $\pm$        2.92   &         2.64   &        39.86   &        0.033   \\
         4   &   587724232647704698   &   02 14 04.4 &+13 11 56.4   &   i   &     0.041659   &        10.40   &        106.4 $\pm$        6.18   &         2.16   &        41.25   &        0.853   \\
         5   &   587724233712205954   &   00 47 30.3 &+15 41 49.5   &   l   &     0.031463   &        11.21   &        128.8 $\pm$        3.08   &         2.59   &        39.94   &       -0.819   \\
         6   &   587724231569309775   &   01 30 06.2 &+13 17 02.1   &   l   &     0.037992   &        10.16   &         36.6 $\pm$       10.60   &         1.90   &        40.69   &        2.153   \\
         7   &   587724234257924238   &   02 11 21.8 &+14 30 15.5   &   l   &     0.041690   &        10.56   &         81.4 $\pm$        5.73   &         2.44   &        40.72   &        0.876   \\
         ...\\
\enddata
\tablenotetext{1}{Galaxy Zoo morphology: e - early-type; l - late-type, i - indeterminate-type. See Section  \ref{sec:gz_morph}}
\tablecomments{Table \ref{tab:cat} is published in its entirety in the 
electronic edition of the {\it Astrophysical Journal}.  A portion is 
shown here for guidance regarding its form and content.}
\label{tab:cat}
\end{deluxetable*}


\subsection{What about Indeterminate-type AGN Host Galaxies?}
Throughout our entire discussion, we have mostly considered the early- and late-type samples as the two morphological extremes. However, both in terms of the entire galaxy sample and the AGN sample, the indeterminate-types are the most numerous class (45.3\% of galaxies and 44.5\% of AGN; see Tables \ref{tab:gzmorph} and \ref{tab:sy_prop}).Future imaging data, such as from PanSTARRS\footnote{The Panoramic Survey Telescope \& Rapid Response System; see \texttt{http://pan-starrs.ifa.hawaii.edu/public/}.}, VST\footnote{The V  LT Survey Telescope ; \texttt{http://www.eso.org/ sci/observing/policies/PublicSurveys/ sciencePublicSurveys.html\#VST}.} or LSST\footnote{The Large Synoptic Survey Telescope; \texttt{http://www.lsst.org/lsst}.}, classified via a future Galaxy Zoo project would enable us to better separate them into early- and late-types.

\subsection{The Black Hole at the Center of the Milky Way should be Active!}
\label{sec:mw}
Figure \ref{fig:sy_frac} shows that the duty cycle for AGN in late-type galaxies is very strongly peaked at a particular stellar mass and color. The Milky Way most likely occupies this sweet spot for late-type AGN. Estimates of the stellar mass of the Milky Way vary, but are around $\sim10^{11}$\Msun, while the current star formation rate is approximately 3 \Msunyr\ \citep[e.g.,][]{1986FCPh...11....1S,2006Natur.439...45D,2009eimw.confE...8C}. These parameters place the Milky Way right where the AGN fraction for late-type galaxies is highest, between 5 and 10\% and therefore the duty cycle is the highest. However, this also implies that the black holes in galaxies like the Milky Way are \textit{not} active $\sim92-95\%$ of the time at the luminosity of the typical Seyfert in the SDSS Universe ($\sim10^{40}$\ergs, see Section \ref{sec:complete} and \ref{sec:sy_prop}).

The black hole at the center of the Milky Way at Sagittarius A$^*$ (Sgr A$^*$) is, apart from occasional weak flares, remarkably quiet with an extremely sub-Eddington accretion rate \citep[e.g.,][]{2006ApJ...640..308M,2008ApJ...682..373M}. However, there is evidence that Sgr A$^*$ was substantially more active in the very recent past. Hard X-ray observations with \textit{ASCA} showed Fe K$\alpha$ lines around Galactic Center emitted by the molecular cloud complex surrounding the black hole. Follow-up observations with more recent hard X-ray facilities indicate that Sgr A$^*$ was orders of magnitude, perhaps $10^6$ times more luminous in the last few hundred years than it is currently. \textit{Suzaku} observations by \cite{2008PASJ...60S.191N} detect a K$\alpha$ luminosity consistent with Sgr A$^*$ having been at $10^{38-39}$\ergs\ a mere 300 yrs ago, while \cite{2004A&A...425L..49R} used \textit{INTEGRAL} data to argue that Sgr A$^*$ was at a 2--200 keV luminosity of $L\sim 1.5 \times 10^{39}$\ergs\ just 300 to 400 years ago. This is not quite the luminosity necessary to excite a narrow-line region sufficiently luminous to be detected by an SDSS fibre at $z \sim 0.02$, but these observations clearly imply that Sgr A$^*$ is active at slightly lower luminosity on short timescales. Together with the duty cycle for Milky Way-like galaxies derived here, these observations may give a first hint of the lifetime of AGN in such galaxies. A caveat to this statement is the fact that the luminosity of this very recent flare is still below the AGN luminosity limit of our sample.

Under what conditions would the black hole at the center of the Milky Way reach a sufficient luminosity so that it would be included in our AGN sample? To reach a typical luminosity of a few percent Eddington $(\sim \rm{few} \times 10^{42}$\ergs\ bolometric luminosity) requires a mass-accretion rate of only $10^{-3}$\Msunyr. This only a factor 100 higher than Bondi accretion rate inferred from simulations, and there are $10^{4}$\Msun\ of molecular gas within 2 pc of the central black hole \citep{2002ApJ...579L..83H}, with large clouds at further distances. It would not necessarily be surprising if dynamical friction or a minor merger, e.g., the swallowing of the Sagittarius dwarf galaxy drove some of these clouds in, temporarily re-igniting the black hole at the center of the Milky Way. All the ingredients thus appear to be on hand for turning our galaxy into a Seyfert AGN. Indeed, the mystery may rather be why is our central black hole so dark and why is gas accretion is so infrequent?

The implication of this work is that the Milky Way is in a location on the color-mass diagram where late-type galaxies have the highest duty cycle for AGN phases. The mass of the Milky Way black hole has been inferred to be $4.1 \times 10^{6}$\Msun\ \citep{2008ApJ...689.1044G} which places it in the typical range of black hole masses in late-type galaxies that are currently growing in the local Universe (Figure \ref{fig:sy_mbh} and \ref{fig:sy_fmbh}). Thus,  5-10\% duty cycle for galaxies like the Milky Way places into context the current quiescence of Sgr A$^*$, together with the evidence for accretion in the recent past and apparently favorable conditions for a resumption of accretion.

\section{Summary}
\label{sec:summary}

 In order to understand the co-evolution of galaxies and the supermassive black holes at their centers, we have investigated which galaxies are more likely to host active black holes and how they differ systematically from their normal counterparts, if at all. We use data from the Sloan Digital Sky Survey and visual classifications of morphology from the Galaxy Zoo project to analyze black hole growth in the nearby Universe ($ z < 0.05$) and dissect the AGN host galaxy population by color, stellar mass and morphology. We show for the first time the importance of host galaxy morphology for black hole growth and relate it to our understanding of the way in which galaxies and their supermassive black holes co-evolve. In summary:

\begin{enumerate}
\item We confirm that the selection of AGN via emission line diagnostic diagrams may miss low-luminosity AGN (\LOIII\ $< 10^{40}$\ergs) in star-forming galaxies.

\item AGN host galaxies as a population have high stellar masses around $10^{11}$\Msun\, reside in the green valley, and have median black hole masses around \Mbh\ $\sim 10^{6.5}$\Msun. 

\item When we divide the AGN host galaxy population into early- and late-types, key differences appear. While both early- and late-type AGN host galaxies have similar typical black hole masses and Eddington ratios, their stellar masses are different, with the early-type hosts being significantly less massive ($\sim10^{10}$\Msun) than the late-type hosts ($\sim10^{11}$\Msun). The late-type hosts nevertheless have the same characteristic bulge masses, and therefore black hole masses, as the early-types. The difference in stellar mass is due to the presence of a massive stellar disk in the late-types.

\item We consider the fraction of galaxies hosting an AGN as a function of black hole mass. Dividing by morphology into early- and late-type galaxies, we find that in early-type galaxies, it is preferentially the galaxies with the \textit{least massive} black holes that are active. In late-type galaxies, it is preferentially the \textit{most massive} black holes that are active, with a potential drop in the active fraction above (\Mbh\ $ \sim10^{7.5}$\Msun). 

\item We estimate the duty cycle of AGN on the color-mass diagram, split by morphology. While some early-type galaxies lie on the red sequence that host AGN, their duty cycle is negligible. The duty cycle of AGN in early-type galaxies is strongly peaked in the green valley below the low-mass end of the red sequence and above the blue cloud. The duty cycle of AGN in late-types, on the other hand, is high in massive green and red late-types. When we impose a minimum Eddington ratio cut, we find that only the green valley early-type AGN still have a substantial duty cycle. No part of the late-type population has a high duty cycle of high-Eddington ratio accretion.

\item We discuss the implications of these results for our understanding of the role of AGN in the evolution of galaxies. We conclude, in particular, that \textit{there are two fundamentally different modes of black hole growth at work in early- and late-type galaxies}. We now have a good understanding of the role of AGN in early-type galaxies: The low-mass early-type AGN hosts are post-starburst objects moving towards the low-mass end of the red sequence. They may represent a low-mass, low-luminosity or \textit{downsized} version of the mode of evolution that was involved in the formation of more massive early-type galaxies at high redshift. The role of AGN in the evolution of the late-type galaxies is less clear and we offer some possible scenarios. The high host stellar masses and grand-design stellar disks makes it implausible that they are the product of recent major mergers moving from the blue cloud to the red sequence. 

\item Finally, the Milky Way likely resides in the `sweet spot' on the color-mass diagram where the duty cycle of AGN is highest amongst late-type galaxies and that the AGN duty cycle for galaxies like the Milky Way is $\sim5-8\%$.
\end{enumerate}

We provide the entire AGN host galaxy sample used in this paper, including derived properties such as stellar masses, black hole masses, $[\mbox{O\,{\sc iii}}]$ luminosities and Eddington parameters, in Table \ref{tab:cat}.

\acknowledgements 

This work would not have been possible without the contributions of citizen scientists as part of the Galaxy Zoo project. We would like to thank Erin Bonning, Timothy Heckman, Guinevere Kauffmann, Taysun Kimm, John Parejko and Sukyoung Yi for helpful discussions.

Support for the work of K.S. was provided by NASA through Einstein Postdoctoral Fellowship grant number PF9-00069 issued by the Chandra X-ray Observatory Center, which is operated by the Smithsonian Astrophysical Observatory for and on behalf of NASA under contract NAS8-03060. K.S. gratefully acknowledges earlier support from Yale University. Support from NSF grant \#AST0407295 and Spitzer JPL Grant \#RSA1288440 is gratefully  acknowledged. S.V. acknowledges support from a graduate research scholarship awarded by the Natural Science and Engineering Research Council of Canada (NSERC). Support for the work of E.T. was provided by the National Aeronautics and Space Administration through Chandra Postdoctoral Fellowship Award Number PF8-90055 issued by the Chandra X-ray Observatory Center, which is operated by the Smithsonian Astrophysical Observatory for and on behalf of the National Aeronautics Space Administration under contract NAS8-03060. S.K. acknowledges support through a Research Fellowship from the Royal Commission for the Exhibition of 1851, a Junior Research Fellowship from Imperial College London and  Senior Research Fellowship from Worcester College, University of Oxford. K.L.M. acknowledges funding from the Peter and Patricia Gruber Foundation as the 2008 Peter and Patricia Gruber Foundation International Astronomical Union Fellow.

Funding for the SDSS and SDSS-II has been provided by the Alfred P. Sloan Foundation, the Participating Institutions, the National Science Foundation, the U.S. Department of Energy, the National
Aeronautics and Space Administration, the Japanese Monbukagakusho, the Max Planck Society, and the Higher Education Funding Council for England. The SDSS Web Site is \texttt{http://www.sdss.org/}.

The SDSS is managed by the Astrophysical Research Consortium for the Participating Institutions. The Participating Institutions are the American Museum of Natural History, Astrophysical Institute Potsdam, University of Basel, University of Cambridge, Case Western Reserve University, University of Chicago, Drexel University, Fermilab, the Institute for Advanced Study, the Japan Participation Group, Johns Hopkins University, the Joint Institute for Nuclear Astrophysics, the Kavli Institute for Particle Astrophysics and Cosmology, the Korean Scientist Group, the Chinese Academy of Sciences (LAMOST), Los Alamos National Laboratory, the Max-Planck-Institute for Astronomy (MPIA), the Max-Planck-Institute for Astrophysics (MPA), New Mexico State University, Ohio State University, University of Pittsburgh, University of Portsmouth, Princeton University, the United States Naval Observatory, and the University of Washington.

This research has made use of NASA's Astrophysics Data System Bibliographic Services. \\
{\it Facilities:} \facility{Sloan()}

\bibliographystyle{hapj}


\newpage
\clearpage

\appendix

\section{Reliability of Stellar Mass Measurements of AGN Host Galaxies}
\label{sec:reliability}

\begin{figure}
\begin{center}

\includegraphics[angle=90, width=0.49\textwidth]{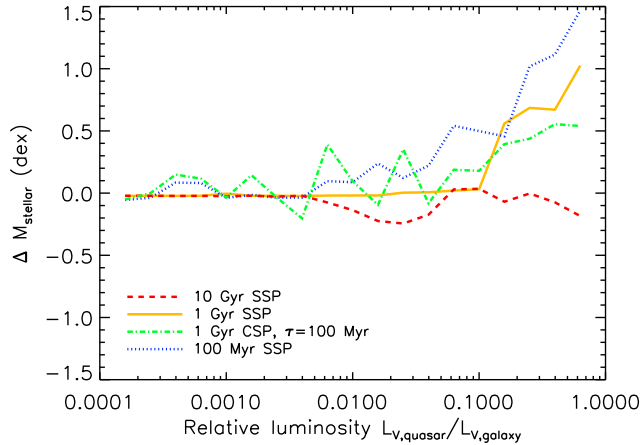}
\caption{The effect of AGN contribution to stellar mass estimates from optical photometric data. We show the results of a test of the effect of adding a quasar template \citep[from][]{2001AJ....122..549V} to an ensemble of galaxy templates varying the relative $V$-band luminosity of the quasar and the galaxy template. This test shows that we are able to recover the stellar mass within 0.2 dex for galaxies at least ten times more luminous in $V$ than the quasar. Stellar mass measurements of host galaxies of brighter quasars degrade as the quasar luminosity approaches the host galaxy.\label{fig:agn_stellarmass}}

\end{center}
\end{figure}

The reliability of galaxy parameters determined from spectral energy distribution fitting techniques is limited by uncertainties in the input models and parameters  \citep[e.g.,][]{2006ApJ...652...85M, 2009ApJ...699..486C, 2009arXiv0904.0002C}. As we perform this stellar mass measurement not only for normal galaxies, but also for the host galaxies of obscured AGN, we need to assess the effect of he presence of a small amount of AGN continuum on stellar mass measurements. We generate a set of model SEDs by combining a galaxy template, $ f_{\lambda, \rm galaxy}$ with a quasar template $f_{\lambda, \rm quasar}$,  \citep[taken from][]{2001AJ....122..549V} and normalising it by a constant $\alpha$ to achieve a specific ratio of $L_{V, \rm quasar}/L_{V, \rm galaxy}$ in order to explore a range of relative luminosities, specified by the constant, $\alpha$:

\begin{equation}
f_{\lambda, \rm model} =  f_{\lambda, \rm galaxy}  + \alpha f_{\lambda, \rm quasar}
\end{equation}

The galaxy templates used are SSPs (single stellar populations) with ages of 100 Myr (young), 1 Gyr (intermediate) and 10 Gyr (old) and a 1 Gyr old exponentially declining CSP (composite stellar population). The results of this experiment are shown in Figure \ref{fig:agn_stellarmass}. We find that for the older galaxy templates, the stellar mass estimates remain relatively unaffected up to a luminosity ratio of $L_{V, \rm quasar}/L_{V, \rm galaxy} \sim 0.1$. The stellar mass estimates for the younger templates begin to become uncertain at the level of $\sim0.2$ dex  when $L_{V, \rm quasar}/L_{V, \rm galaxy}$ reaches $\sim0.01$, which is likely the maximum amount of contamination that would be expected in our sample.

Adding the quasar template to the galaxy is comparable to adding a very young stellar population. This explains the perhaps counterintuitive outcome of old stellar populations being relatively unaffected - the two-component code identifies the quasar as a small, very young stellar population with a very low mass-to-light ratio and isolates it from the bulk old stellar population. The SED fitting method has more trouble with galaxy templates where young stellar populations are genuinely present. In this case, the quasar template becomes entangled with the young population and so leads the code to identify a young stellar population of the wrong age and mass-fraction, resulting in a greater uncertainty in the measured stellar mass. This means that this effect is not a weakness of two-component codes, but a generic problem for all methods.

This experiment implies that the uncertainty introduced by AGN is comparable to the other uncertainties that affect stellar mass measurements. The AGN in the relatively local volume probed by our sample are low-luminosity and obscured, so their expected AGN continuum contribution is small. As a consequence, the stellar mass estimates of AGN host galaxies are approximately as reliable as those of normal galaxies.

\section{Assessment of the Impact of Low-luminosity AGN and LINERs}
\label{sec:liners}

\begin{figure*}[!ht]
\begin{center}

\includegraphics[angle=90, width=\textwidth]{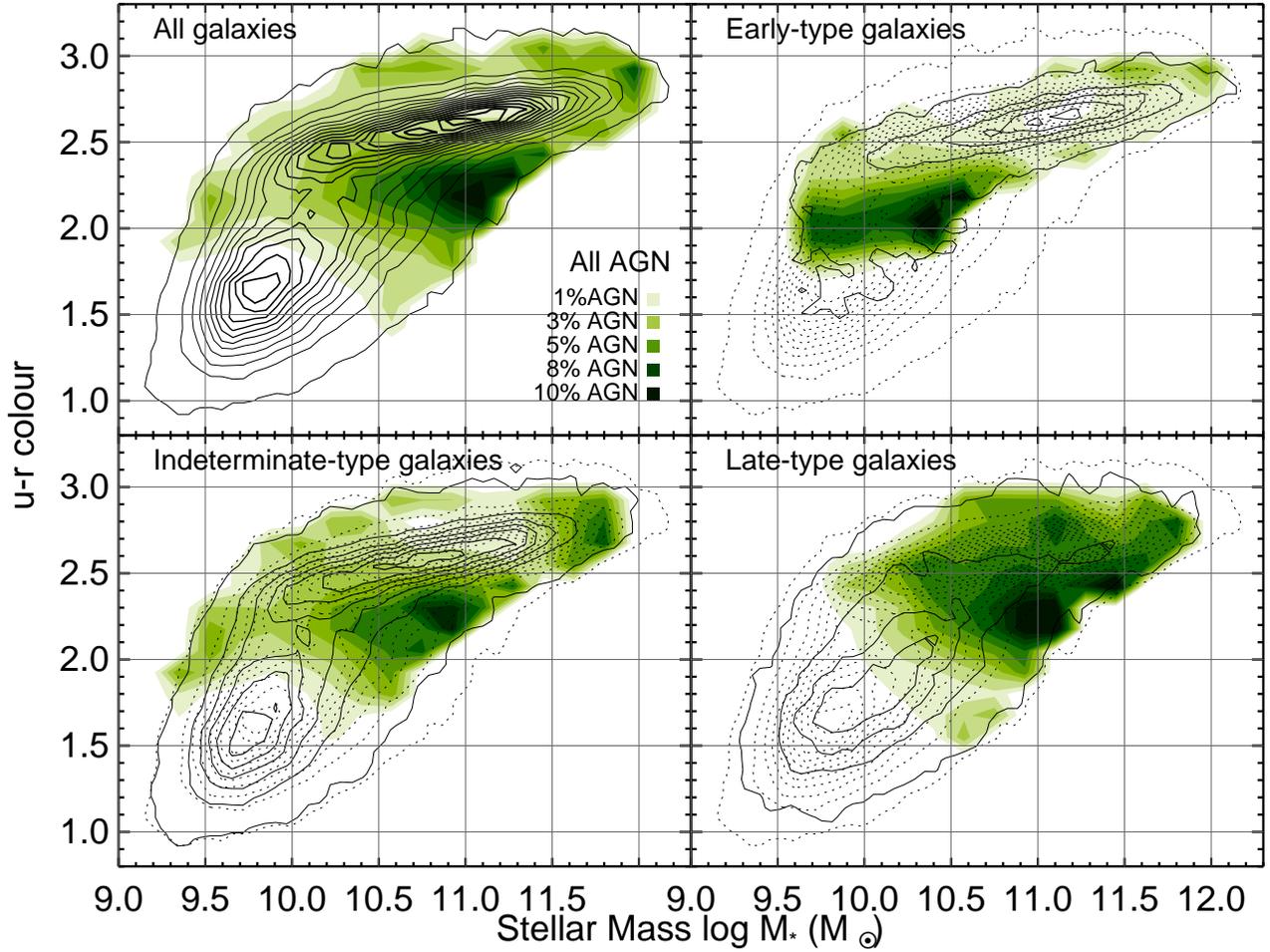}
\caption{Same are Figure \ref{fig:sy_frac}, but for all objects classified as Seyfert or LINER with \LOIII $> 10^{40}$\ergs. This Figure illustrates that the inclusion of the small fraction of LINERs that have high \OIII\ luminosities does not substantially change the results presented in this work. }

\end{center}
\end{figure*}

Since LINER spectra are preferentially detected in massive red sequence early-type galaxies \citep[e.g.,][]{2006MNRAS.372..961K,2007MNRAS.382.1415S}, the exclusion of LINERs may bias our AGN host galaxies sample. If LINERs are AGN, then their implied Eddington ratios inferred from \LOIIIMbh\ are systematically lower than those of Seyferts \citep{2006MNRAS.372..961K}. This implies that their \OIII\ luminosities are substantially lower for a given black hole mass, which in turn means that the tpyical LINER \OIII\ luminosity is significantly lower than that of Seyferts. In fact, only 19\% of all LINERs detected using our BPT diagrams have \OIII\ luminosities greater than the completeness limit of  $\sim 10^{40}$\ergs that we determine in Section \ref{sec:complete}. In order to test whether our removal of LINERs from our sample significantly affects our results, we perform the following experiment: we include all LINERs in the AGN sample, but remove all objects (whether classified Seyfert or LINER) with \LOIII $< 10^{40}$\ergs\ (this removes $\sim$20\% of the Seyfert sample as well). When we then redo the analysis of this paper using this new sample, we find that all the qualitative trends we describe remain unchanged, and that the quantitative differences are small. 

\end{document}